\documentclass[11pt,a4paper]{article}
\pdfoutput=1
\usepackage{graphicx}
\usepackage[tbtags]{amsmath}
\usepackage{amssymb}
\usepackage{amsthm}
\usepackage{array}
\usepackage{xy}
\usepackage{slashed}

\usepackage[iso-8859-7]{inputenc}

\usepackage{graphicx}

\def\<{\left\langle}
\def\>{\right\rangle}

\title{Reheating predictions in gravity theories with derivative coupling  \normalfont}

\author{\small {Ioannis Dalianis, George Koutsoumbas, Konstantinos Ntrekis, Eleftherios Papantonopoulos} \\
\small{ Department of
Physics, National Technical University of Athens,} \\
\small {Zografou Campus
GR 157 73, Athens, Greece}}
\begin{document}


\newcommand{\g}{\greektext}
\newcommand{\e}{\latintext}
\maketitle
\abstract{ We investigate the inflationary predictions of a simple Horndeski theory where the inflaton scalar field has a non-minimal derivative coupling  (NMDC) to the Einstein tensor. The NMDC is very motivated for the construction of successful models for inflation, nevertheless its inflationary predictions are not observationally distinct.
We show that it is possible to probe the effects of the NMDC on the CMB observables by taking into account both the dynamics of the inflationary slow-roll phase and the subsequent reheating.
We perform a comparative study between representative inflationary models with canonical fields minimally coupled to gravity and models with NMDC. We find that the inflation models with dominant NMDC generically predict a higher reheating temperature and a different range for  the tilt of the scalar perturbation spectrum $n_s$ and scalar-to-tensor ratio $r$, potentially testable by current and future CMB experiments.}
\\
\\
\tableofcontents

\section{Introduction}

Scalar-tensor theories are non-trivial extensions   of the  Einstein gravity
giving interesting modifications to the standard theory in both small and large distances.
Among them  there is a gravity theory which includes a term of  a direct coupling of a scalar field to Einstein tensor.
This term belongs to a general class of
scalar-tensor gravity theories resulting from the Horndeski
Lagrangian \cite{Horndeski:1974wa}. These theories, which were
recently rediscovered \cite{Deffayet:2011gz},
 give second-order field equations and contain as a subset a theory
which preserves classical Galilean symmetry in flat spacetime \cite{Nicolis:2008in,
Deffayet:2009wt, Deffayet:2009mn, Kobayashi:2011nu}.
The simplest of the Horndeski theories has an action which apart from the minimal coupling of the scalar
field to gravity includes a derivative coupling to Einstein tensor.  
The Lagrangian reads,
\begin{equation} \label{NMDC1}
{\cal L}=\sqrt{-g} \left[\frac12 M^2_P R -\frac12\left(g^{\mu\nu} -\frac{G^{\mu\nu}}{\tilde{M}^2} \right)\partial_{\mu}\phi\partial_{\nu}\phi - V_\text{}(\phi)\right]\,.
\end{equation}
As can be seen in the above action, the derivative coupling of the scalar field to Einstein tensor
introduces a new scale $\tilde{M}$ in the theory
which has interesting implications on short distances for the black hole physics  \cite{Kolyvaris:2011fk, Rinaldi:2012vy, Kolyvaris:2013zfa, Koutsoumbas:2015ekk} and on large distances for inflation \cite{Amendola:1993uh}.

The rather attractive feature for the inflation model building point of view is that the derivative coupling acts as a friction mechanism assisting in the realization of the slow-roll phase even when the potential slope is too steep \cite{Amendola:1993uh,Sushkov:2009hk, Germani:2010gm}. Potentials that fail to drive an accelerating expansion become viable candidates for inflation when the Non-Minimal-Derivative-Coupling (NMDC) is introduced. Some very interesting examples of 
 such non-inflating potentials in General Relativity (GR)
are the Standard Model Higgs potential $V\propto \lambda \phi^4$ that has a selfcoupling $\lambda\sim 0.1$ \cite{Germani:2010gm},
the pseudo-Nambu-Goldstone potential, $V\propto (1-\cos(\phi/f))$, with a typical scale of variation $f$ lying below the Planck mass, $M_\text{Pl}$ \cite{Germani:2010hd}, and the exponential potential, $V\propto e^{-\lambda_e \phi/M
_\text{Pl}}$ with $\lambda_e \gg 1$ \cite{Tsujikawa:2012mk, Dalianis:2014nwa}. Furthermore, prototype inflationary models such as the quadratic and the (generic) quartic potentials, which are currently mostly or fully excluded, can successfully fit the last observational data thanks to the NMDC which yields a suppressed tensor-to-scalar ratio.
In addition, the presence of the NMDC does not enhance the non-Gaussian fluctuations of the scalar field  albeit, are found to vanish at order of the fist slow-roll parameter \cite{Germani:2011ua}.
Observational constraints on a number of representative inflationary models with a field
derivative coupling to the Einstein tensor can be found in Ref. \cite{Tsujikawa:2012mk, Tsujikawa:2013ila}.

The NMDC effects are suppressed by the mass scale $\tilde{M}$ and become manifest when the Hubble energy scale, $H$, is much larger. Namely, when $H_\text{} \gg \tilde{M}$ the NMDC rules the evolution of the non-minimally coupled scalar and modifies the inflationary dynamics whereas, in the opposite case, $H_\text{} \ll \tilde{M}$, the non-minimal coupled-scalar behaves approximately as a standard canonical field.

Although the new mass scale $\tilde{M}$ of the non-minimal derivative coupling is usually considered to be well below the Planck mass, the non-minimally coupled graviton-inflaton system remains in the weak coupling regime during inflation. Actually, for a slowly rolling scalar field the strong coupling scale of the derivative coupling theory in a homogeneous and isotropic background is found to be around $M_\text{Pl}$ \cite{Germani:2010gm, Germani:2010hd,Germani:2011bc, Germani:2011ua, Germani:2014hqa}. The quantum corrections are also found to be suppressed due to fact that the theory possesses an asymptotic local shift symmetry when the non-minimal derivative coupling dominates \cite{Germani:2011bc}, see also \cite{Kamada:2010qe, Kunimitsu:2015faa} for relevant generalized set-ups.  Remarkably that this higher derivative theory has been supersymmetrized in the framework of the new-minimal supergravity \cite{Farakos:2012je} and inflationary potentials can be consistently generated via the gauge-kinematic function \cite{Dalianis:2014sqa}. The thermodynamics of general scalar-tensor theory with NMDC was discussed in \cite{Huang:2016qkd}.

The NMDC is well motivated from the theoretical and the observational sides, however a decisive observational test for the presence of the NMDC in the dynamics of the early Universe remains challenging. Needless to say, that fitting the data does not account for an automatic selection of a model. There are GR inflationary models with canonical fields characterized by similar inflationary dynamics and predictions. Indeed, it has been shown that during the (quasi) de-Sitter phase the NMDC inflaton system enjoys a dual description in terms of a minimally coupled field with GR gravity \cite{Germani:2011mx, Dalianis:2014sqa}. This ascertainment suggests
 that, at first sight, cosmological models with NMDC for the scalar field to Einstein tensor cannot give
a distinct feature compared to cosmological models with a canonical coupling and GR gravity. Therefore, the inflationary dynamics can not differentiate these cosmological models. A clear difference appears however after the end of inflation.

The energy stored in the omnipresent inflaton, which is expected to coherently oscillate about its minimum after inflation, has to be converted to a plasma of relativistic particles 
and this intermediate transition phase is called reheating \cite{Kofman:1997yn}. The presence of the NMDC can modify the standard picture of the reheating phase. When the NMDC is active the inflaton oscillates rapidly without significant damping \cite{Sadjadi:2012zp, Ghalee:2013ada, Sadjadi:2013na, Gumjudpai:2015vio, Jinno:2013fka, Ema:2015oaa, Myung:2016twf, Ema:2016hlw}.
The Hubble parameter also oscillates and the expansion law is significantly different than the standard case of a canonically coupled inflaton \cite{Sadjadi:2012zp, Ghalee:2013ada, Jinno:2013fka, Ema:2015oaa}.
The implication of such oscillations, where the NMDC dominates over the canonical kinetic term, may be essential 
for the stability of the post-inflationary system since the sound speed squared of short-wavelength scalar perturbations oscillates between positive and negative values \cite{Ema:2015oaa}.
In Ref. \cite{Germani:2015plv} it was shown that for the new-Higgs inflation the linear regime is actually well possible not to be violated during the reheating period,  despite the amplification of the Newtonian potentials, and the predictions of NMDC $\phi^4$-inflation models to remain valid. We also mention that the exponential inflationary potentials are free form the instabilities issues because the reheating period starts only after the NMDC term becomes subdominant. 
Another difference with the canonical kinetic coupling was discussed in
\cite{Koutsoumbas:2013boa}. It was found that there is a suppression of heavy particle production in the preheating period after
inflation, as the derivative coupling gets increased. This was attributed to the fast decrease of  kinetic
energy of the scalar field due to its  wild oscillations.

The above discussion suggests that
the reheating period may differentiate the predictions of cosmological models with NMDC compared to models with the canonical coupling.
Observationally, there are no cosmological observables directly related with the reheating period.
There is nevertheless the possibility to trace back the cosmological evolution for observable CMB scales from the time of the first  Hubble crossing to the present time and, in this way, to constrain the reheating period. For a given model, the duration and final temperature of the reheating period, as well as the mean reheating equation of state (EoS), can be directly  linked to inflationary observables \cite{Martin:2010kz, Mielczarek:2010ag, Bezrukov:2011gp, Martin:2014nya, Munoz:2014eqa, Dai:2014jja, Gong:2015qha, Cook:2015vqa, Rehagen:2015zma, Drewes:2015coa}.

In this work we show that it is in principle possible to observationally probe the effects of the NMDC
by taking into account both the dynamics of the inflationary quasi-de-Sitter phase and the subsequent reheating.
Neglecting the reheating stage, the models ($\phi, V_\text{}(\phi)$) with dominant NMDC give inflationary predictions similar to those of models ($\varphi, V_\text{GR}(\varphi)$) where $\varphi$ a canonical scalar with GR, see Table 1.
This degeneracy would render impossible a selective test for the inflationary NMDC models, if the expansion rate during the reheating stage was not much different. We find that the peculiar effective EoS value of the NMDC models yields different e-folds number and a dramatically different reheating temperature, providing an important evidence
for the model selection analysis: {\itshape the range for  the tilt of the scalar perturbation spectrum $n_s$ and ratio $r$ of
scalar-to-tensor perturbation amplitudes differ when the NMDC dominates over the canonical term during the reheating stage}. For a given temperature our results indicate a simple check that can support, disfavor or 
leave open the existence of the NMDC by employing the  measurements of the CMB anisotropies -under the assumption of simple early cosmic evolution. A general conclusive check is not possible by the current {\itshape Planck} satellite measurements \cite{Ade:2015xua, Ade:2015lrj}, though the parameter space is much constrained for specific models, such as the exponential.   Forthcoming measurements of $n_s$ and $r$ by future experiments EUCLID \cite{Amendola:2012ys}, PRISM \cite{Andre:2013afa} and LiteBIRD \cite{Matsumura:2013aja}  should be able to diminish further the observational uncertainties and test more accurately the distinct NMDC inflationary predictions. In addition, observational constraints on the reheating temperature, as those proposed by the DECIGO \cite{Kuroyanagi:2014qza}, will be of decisive importance. On the theoretical, the validity of the the perturbation theory  during the postinflationary stage for the NMDC models should be thoroughly investigated along the lines of Ref. \cite{Ema:2015oaa, Germani:2015plv, Myung:2016twf}.

The paper is organized as follows.  In section 2, we outline the connection between the inflationary observables and the expansion history of the universe. In section 3, we present standard results of the GR monomial inflationary models in order later to make manifest the differences with the NMDC. 
In Section 4, we overview the dynamics of an inflaton with NMDC and the degeneracy in the predictions with GR models. In section 5, we turn to the inflationary predictions for the  NMDC and we derive the formulas for e-folds number of expansion.
In section 6, we discuss the observational consequences and compare the full predictions of the NMDC with their dual GR models. In section 7, we conclude and discuss prospects for further research. Finally, in the appendix  a survey for the correspondence between NMDC and GR models can be found.

\begin{table}
\centering
\begin{tabular}{|  p{0.44\textwidth} | p{0.36\textwidth} |}
\hline
\hline
\quad\quad Inflaton with NMDC & \quad \quad Minimally coupled inflaton
\\[0.5ex]
\hline
$\frac12 M^2_P R -\frac12\left(g^{\mu\nu} -\frac{G^{\mu\nu}}{\tilde{M}^2} \right)\partial_{\mu}\phi\partial_{\nu}\phi - m^2_\phi \phi^2/2$ &  $ \frac12 M^2_P R -\frac12\partial^{\nu}\varphi\partial_{\nu}\varphi -\mu^3_\varphi \varphi$
\\[0.5ex]
\hline
$\frac12 M^2_P R -\frac12\left(g^{\mu\nu} -\frac{G^{\mu\nu}}{\tilde{M}^2} \right)\partial_{\mu}\phi\partial_{\nu}\phi - \lambda_\phi \phi^4$ &  $ \frac12 M^2_P R -\frac12\partial^{\nu}\varphi\partial_{\nu}\varphi - \xi^{8/3}_\varphi \varphi^{4/3}$
\\[0.5ex]
\hline
$\frac12 M^2_P R -\frac12\left(g^{\mu\nu} -\frac{G^{\mu\nu}}{\tilde{M}^2} \right)\partial_{\mu}\phi\partial_{\nu}\phi - V_0e^{-{\lambda_e} \phi/{M_P}}$ &  $ \frac12 M^2_P R -\frac12\partial^{\nu}\varphi\partial_{\nu}\varphi -m^2_\varphi \varphi^{2}/2$
\\[0.5ex]
\hline
\hline
\end{tabular}
\caption{The correspondence between models during the slow-roll phase. The relation between the mass parameters of the two theories can be found in the appendix.}
\end{table}

Throughout the paper the symbol $\varphi$ stands for an inflaton scalar field with canonical kinetic term and GR and the $\phi$ for an inflaton scalar field with NMDC.


\section{Number of e-folds and observables}

In the inflationary paradigm an observed scale at the CMB was inside the Hubble radius, $r_H=1/H(t)$, during inflation. At the time $t_{\text{*}}$ it exits the inflationary horizon $1/H(t_*)$ and re-enters after inflation at $t_\text{cmb}$. In order to estimate the time $t_*$
we have to exactly know how much the scale is stretched during the postinflationary evolution.  This is encoded in number of e-folds $N$ \cite{Liddle:2003as}.

The main uncertainty comes from the phase between the end of inflation and the start of the radiation era.
The size of given scale $k^{-1}$ that exited the Hubble radius $H^{-1}_k$ during inflation can be related to the size of the present Hubble radius $H^{-1}_0$ via the relation  
\begin{equation}
\frac{k}{a_{0} H_{0}} = \frac{a_k}{a_\text{end}} \frac{a_\text{end}}{a_\text{reh}} \frac{a_\text{reh}}{a_0}\frac{H_k}{H_0}\,,
\end{equation}
where the subscripts refer to the time of horizon crossing ($k$), the time inflation ends (end), the time the reheating stage ends (reh), the epoch of the radiation-matter equality (eq) and the present time (0).
We call $N_k$ the number of e-folds that take place from the time $H_k^{-1}$ till the end of inflation, and $N_\text{reh}$ the  number of e-folds from the end of inflation until the end of reheating,
\begin{equation} \label{Ns}
N_k \equiv \ln \left( \frac{a_\text{end}}{a_k} \right)\,, \quad\quad\, N_\text{reh} \equiv \ln \left(\frac{a_\text{reh}}{a_\text{end}} \right) \equiv \frac{1}{3(1+\bar{w}_\text{reh})} \ln \frac{\rho_\text{end}}{\rho_\text{reh}}\,.
\end{equation}
The symbol $\bar{w}_\text{reh}$ stands for the effective average equation of state (EoS) during the whole reheating period \cite{Martin:2014nya}
\begin{equation} \label{avEoS}
\bar{w}_\text{reh} =\frac{1}{N_\text{reh}}\int^{N_\text{reh}}_{} w_\text{reh}(N)dN\,=\, \frac{1}{\ln\left( \frac{a_\text{reh}}{a_\text{end}}\right)} \int^{a_\text{reh}}_{a_\text{end}} w_\text{reh}(a)\frac{da}{a}\,,
\end{equation}
where $dN=Hdt$. Integrating over the whole reheating period, and not over a single oscillation period, allows to encompass a possible peculiar evolution of the $w(t)$ such as the one found in the NMDC theories.
The energy conservation equation,
$\dot{\rho}_\phi=-3H(\rho_\phi+p_\phi)=3H\rho_\phi(1+w(t))$, together with the $\bar{w}_\text{reh}$ definition above yields the averaged evolution of the energy density $\bar{\rho}(a)=\rho_\text{end}(a/a_\text{end})^{-3(1+\bar{w}_\text{reh})}$ and thus the Eq. (\ref{Ns}) during the reheating stage. The $N_k$ and the $N_\text{reh}$ can be expressed in terms of the measured quantities. Adopting the {\itshape Planck} collaboration pivot scale, $k=k_*=0.05 \text{Mpc}^{-1}$, we get $H_k= H_*=\pi M_\text{Pl}(rA_s)^{1/2}/\sqrt{2}$ where $\ln(10^{10}A_s)=3.089$ from {\itshape Planck} \cite{Ade:2015xua, Ade:2015lrj}. Also the present CMB temperature $T_0=2.72$ K is related to the maximum temperature of the radiation dominated era, $\rho_\text{reh}=(\pi^2/30)g_\text{reh}T^4_\text{reh}$, via the relation $T_\text{reh}=T_0(43/11g_s)^{1/3}a_0 / a_\text{reh}$, see e.g. Ref. \cite{Dai:2014jja, Cook:2015vqa}. In particular, one finds 
\begin{equation}
\frac{1-3\bar{w}_\text{reh}}{4} N_\text{reh} = -N_*+\ln \frac{H_*}{\rho^{1/4}_\text{end}} +C_{*}
\end{equation}
where, $N_*\equiv N_k$ and
\begin{equation}
C_* = -\ln \frac{k_*}{a_0T_0}-\frac14 \ln\frac{30}{g_\text{reh} \pi^2}-\frac13 \ln\frac{11g_\text{reh}}{43}\,.
\end{equation}
Utilizing the slow-roll approximation for the Friedmann equation $H^2_*\simeq V_*/(3M^2_\text{Pl})$ and the relation $r_*=16\epsilon_*$  we can write $\ln (H_*/\rho^{1/4}_\text{end})=1/4\ln(8\pi^2A_s/3)+1/4 \ln \epsilon_* +1/4 \ln (V_*/\rho_\text{end})$. After substituting numbers for $A_s$ and the (nearly) model independent quantity $C_* \simeq 61.7$, we get
\begin{equation} \label{Nreh1}
\frac{1-3\bar{w}_\text{reh}}{4} N_\text{reh} \,\simeq \,  57.5 - N_* + \frac14 \ln \epsilon_* +\frac14 \ln\frac{V_*}{\rho_\text{end}}\,.
\end{equation}
For a given model one can write the ratio $V_*/\rho_\text{end}$ and the slow-roll parameter, $\epsilon$, in terms of $N_*$, and in turn the $N_*$ in terms of the spectral index of the scalar perturbations, the measured quantity $n_s$, $N_*=N_*(n_s)$. Therefore we get the number of e-folds for the reheating phase in terms of $n_s$, $N_\text{reh}=N_\text{reh}(n_s)$, modulo an uncertainty at the $\bar{w}_\text{reh}$ value. Notice that for $\bar{w}_\text{reh}=1/3$ the $N_*$ does not depend on the duration of the reheating phase. In such a case the reheating is either instantaneous or the expansion rate mimics that of a radiation dominated universe. For $\bar{w}_\text{reh}=1/3$ the $N_*$ gets maximized unless values $\bar{w}_\text{reh}>1/3$ are possible. Inflation with NMDC and exponential potential, (\ref{potDC}), is such an example, thought the $\bar{w}_\text{reh}$ value gets suppressed, e.g due to the unavoidable presence of inflationary gravitational waves \cite{Dalianis:2014nwa}.

In addition we can estimate the reheating temperature,
\begin{equation} \label{Treh1}
T_\text{reh}(n_s, \bar{w}_\text{reh})\, = \, \rho_\text{end}^{1/4} \left(\frac{30}{\pi^2 g_*}\right)^{1/4} e^{-\frac{3}{4}(1+\bar{w}_\text{reh})N_\text{reh}(n_s)}\,.
\end{equation}
The $\rho_\text{end}$ is a $N_*$ independent quantity and it can be found from the condition $\epsilon=1$. We can define $V_\text{end}=\gamma \rho_\text{end}$ where, roughly, $\gamma=2/3$. If the inflaton field does not violate the null energy condition, $w_\text{reh}\geq -1$, the expression in front of the exponential at Eq. (\ref{Treh1}) represents the maximum possible temperature, $T_\text{max}$, of the radiation dominated era,
\begin{equation} \label{}
T_\text{reh}(n_s, \bar{w}_\text{reh})\, = \, T_\text{max} \, e^{-\frac{3}{4}(1+\bar{w}_\text{reh})N_\text{reh}(n_s)}\,.
\end{equation}
The $T_\text{max}$ accounts for the reheating temperature when the inflaton field decays immediately after the end of the inflationary era.

For a particular inflation model and a given $\bar{w}_\text{reh}$ the measurement of the spectral index value can determine the duration of the reheating period and the reheating temperature of the universe. The current highest experimental sensitivity in the value of the $n_s$, given by the {\itshape Planck} mission, is a little better than $10^{-2}$.  Future experiments such as EUCLID \cite{Amendola:2012ys}, PRISM \cite{Andre:2013afa} are designed to go down to a $10^{-3}$ error, making a more conclusive model selection possible.

Below we fast overview some standard GR results in order to make the differences with the NMDC manifest.


\section{Inflationary observables for a canonical inflaton with GR}

A simple and generic class of inflationary models is characterized by
a single field monomial potential of the form
\begin{equation} \label{GRpot}
V_\text{}(\varphi) = \lambda_q\, M^4_\text{Pl} \left(\frac{\varphi}{M_\text{Pl}} \right)^q\,.
\end{equation}
This class includes the archetype large field inflationary model, the $\varphi^2$, as well as the linear potential $\varphi$ and fractional powers such as $\varphi^{4/3}$ and $\varphi^{2/3}$. These last models, that are not quadratic, suggest an interesting UV mechanism, such as the axion monodromy inflation \cite{McAllister:2008hb, McAllister:2014mpa}.  In addition, there is a remarkable correspondence with the inflationary models $\phi^p$ with NMDC and $p$ an integer power in the potential.
The expression (\ref{GRpot}) should be seen as the leading potential term during the slow-roll phase and corrections should be considered for lower field values. We will consider that after inflation with potential $V(\varphi)\propto \varphi^q$ where $q=1, 2/3, 4/3$ the inflation field oscillates about a minimum that can be approximated by a smooth potential, such as the quadratic or the quartic, see e.g. Ref. \cite{McAllister:2008hb, McAllister:2014mpa, Adshead:2015pva}.

Inflation takes place for transplanckian values when
the first two slow-roll parameters
\begin{equation} \label{eN}
\epsilon_{V} \equiv \frac{M^2_\text{Pl}}{2} \left(\frac{V'}{V}\right)^2 
\,,\quad\quad\quad \eta_{V} \equiv {M^2_\text{Pl}}\frac{V''}{V} \,,
\end{equation}
are smaller than one.
From the definition of the number of e-folds, $N_*\equiv \int^{t_{end}}_{t_*} H dt$, one finds that
the inflationary predictions $1-n_s=6\epsilon_V-2\eta_V$ and $r=16 \epsilon_V$, to lowest order in slow-roll parameters, in terms of the e-folds number read
\begin{equation} \label{A}
1-n_s \simeq \frac{2q+4}{4N_*+q}\,, \quad\quad r \simeq \frac{16\, q}{4N_* + q}\,.
\end{equation}

\begin{figure} \label{f1}
\centering
\begin{tabular}{cc}
{(a)} \includegraphics [scale=.52, angle=0]{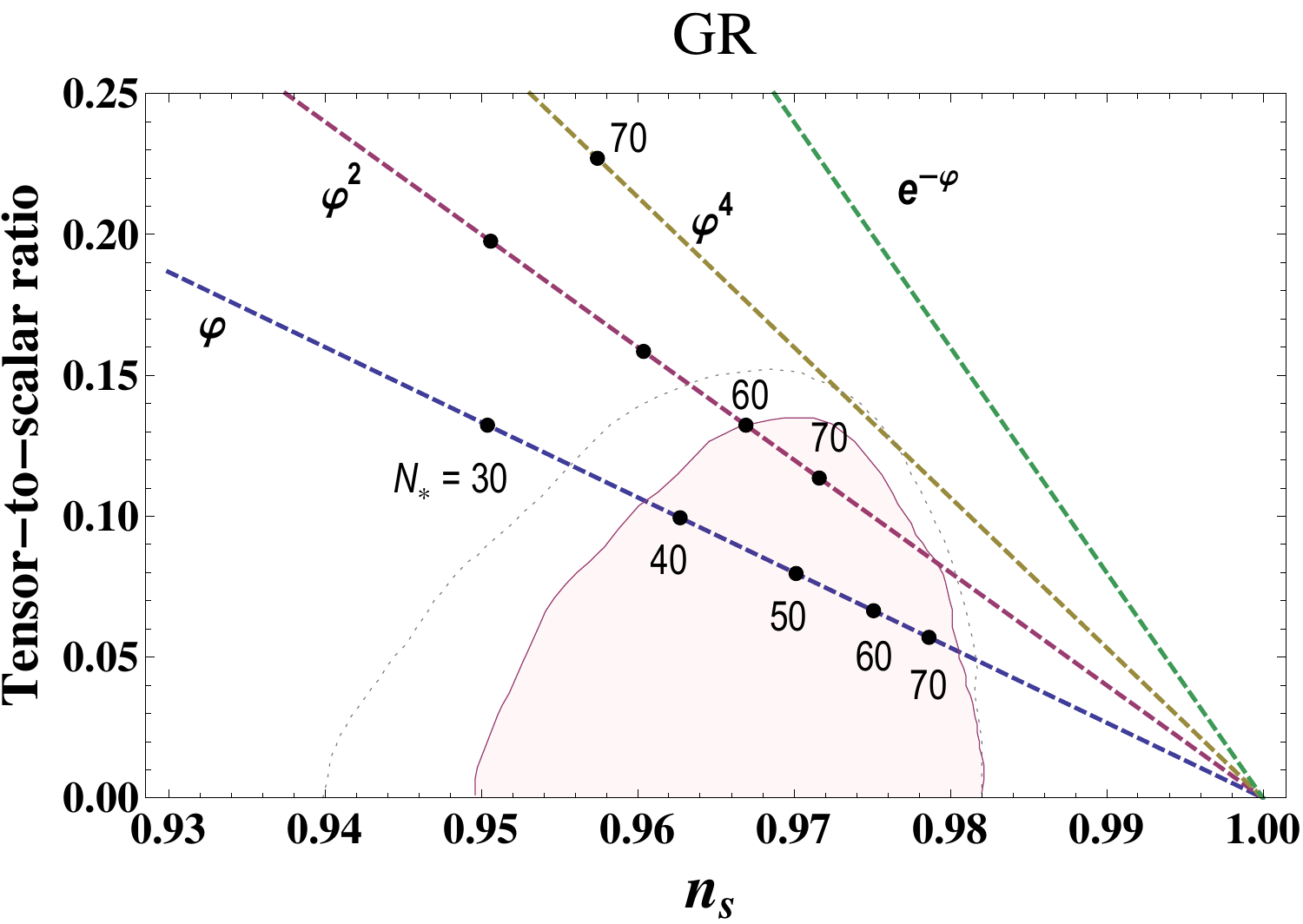} \quad
{(b)} \includegraphics [scale=.52, angle=0]{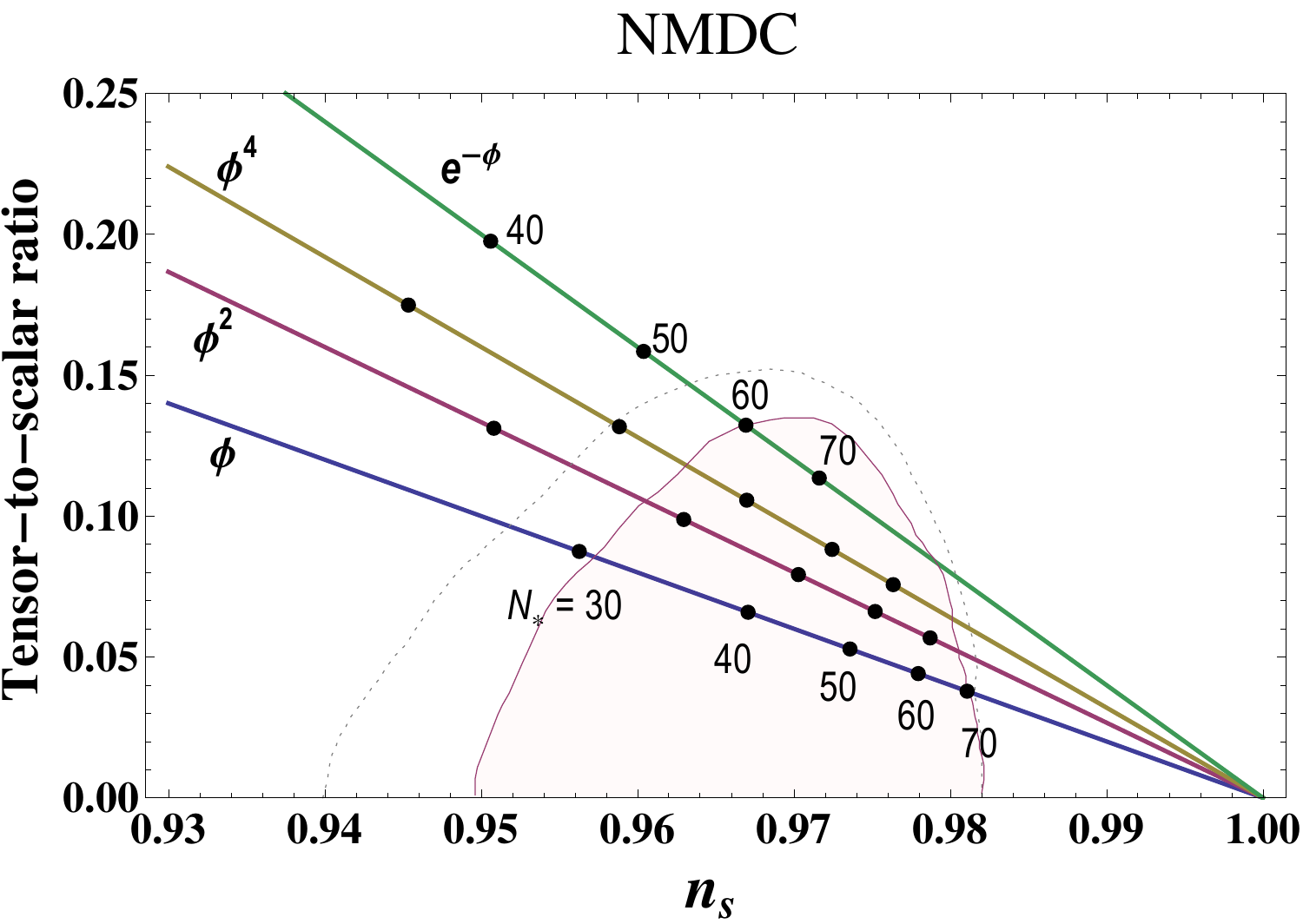}  \\
\end{tabular}
\caption{\small{ The marginalized joint $68\%$ CL regions for the primordial tilt, $n_s$ and the tensor-to-scalar ratio, $r_{0.002}$ from {\itshape Planck} $2013$ (dotted) the from {\itshape Planck} TT$\_\text{lowP}$ $2015$ (the colored contour region).
The left panel shows the theoretical predictions $r=r(n_s)$ for a canonical scalar inflaton with Einstein gravity and $V_\text{}(\varphi) \propto \varphi, \, \varphi^2, \, \varphi^4, \, e^{-\varphi}$. The right shows the theoretical prediction $r=r(n_s)$ for an inflaton with NMDC coupling and $V_\text{}(\phi) \propto \phi, \, \phi^2, \, \phi^4, \, e^{-\phi}$. The black dots show the number of e-folds $N_*$ from $30$ to $70$. Apparently, the models with NMDC fit better the data.
}}
\end{figure}

\subsection{CMB normalization}

The observed amplitude of the CMB anisotropies normalize the coefficient $\lambda_q M^{4-q}_\text{Pl}$ of the potential $V_\text{}(\varphi)$.
The power spectrum of the scalar perturbation, in terms of the gauge invariant $\zeta$ curvature perturbation, reads
\begin{equation} \label{Power}
{\cal P}_\zeta =\frac{H^2_*}{8\pi^2 \epsilon_{V*} M^2_\text{Pl}}\,.
\end{equation}
The ${\cal P}_\zeta$ is an observable and its value is measured at the CMB scale $k^{-1}_*$ by the {\itshape Planck} satellite to be ${\cal P}_\zeta =2\times 10^{-9}$.
Using the slow roll approximation 
the relation for the $\lambda_q$ in terms of the number of e-folds is obtained,
 \begin{equation} \label{gener}
\lambda_q(N_*) = 12 \pi^2 \, q^2 \, \left[\frac{q}{2}(4N_*+q) \right]^{-q/2-1} {\cal P}_\zeta\,.
\end{equation}
In the case of the linear, quadratic and quartic potentials we get respectively $\mu^3_\varphi \sim 2.4 \times 10^{-10} M^3_\text{Pl}$, 
$m_\varphi \simeq  6.9 \times 10^{-6}  M_\text{Pl}$ and  $\lambda_\varphi\, \simeq\, 6 \times 10^{-14}$ for $N_*=50$.

We will see in the section 5 that when one departs from the minimal case of canonical scalar field with Einstein Gravity, such as Horndeski theories, the CMB normalization of the potential is modulated due to the new interactions involved.

\subsection{The reheating stage}

After the end of the acceleration phase the inflaton is expected to oscillate about the minimum of its potential, though run-away scenarios are also possible, gradually transforming its energy to other degrees of freedom. The averaged effective equation of state for a coherently oscillating scalar field is given by the well-known relation \cite{Shtanov:1994ce}
\begin{equation} \label{potw1}
\bar{w}_\text{reh(GR)} \equiv \frac{\left\langle p\right\rangle}{\left\langle \rho\right\rangle}=\frac{q-2}{q+2}\,,
\end{equation}
where $q$ is the power of the monomial potential.
The reheating period leads to the thermalization of the universe, possibly via different distinct stages such as the preheating.
It last for a time $\Gamma_\varphi^{-1}$ where $\Gamma_\varphi$ is the inflaton decay rate. The complete decay of the inflaton signals the end of the reheating period and the initiation of the radiation era with $T_\text{reh}\sim ({\Gamma_\varphi M_\text{Pl}})^{1/2}$. Actually, the process by which the universe reheats is poorly constrained observationally.
There is a vast literature on the subject, see Ref. \cite{Amin:2014eta} for a recent review.
The postinflation value of the effective EoS ranges in the interval $-1/3\leq\bar{w}_\text{reh(GR)}\leq 1$.
Values $\bar{w}_\text{reh(GR)}>1/3$ are difficult to conceive since they require a potential dominated by higher dimensional operators or fractional powers of the potential near the minimum unnatural from a quantum-field theoretical point of view, while values $\bar{w}_\text{reh(GR)}<0$ are not expected since the potentials with fractional powers are approximated by smooth functions about the minimum.
We consider that the shape of the low energy potential (\ref{potw1}) resembles either $\varphi^2$ or $\varphi^4$ and hence we will take
as benchmark values the $\bar{w}_\text{reh(GR)}=0$ or $1/3$. In addition we will take the $\bar{w}_\text{reh(GR)} =1/5$ as benchmark value indicated by numerical studies of thermalization scenarios \cite{Podolsky:2005bw}. Other sample values will be considered in the plots, see Fig. 7 and 8.
We mention that the value $\bar{w}_\text{reh}=1/3$ is equivalent, in terms of the expansion history, with the instantaneous reheating of the universe.

\subsection {The $N_\text{reh}$ and $T_\text{reh}$ in the monomial GR models}
The energy density at the end of inflation can be written as $\rho_\text{end} \simeq \gamma^{-1} V(\varphi_\text{end})\equiv \gamma^{-1} V_\text{end}$, where the $\gamma$ parameter represents the contribution of the $V_\text{end}$ at the total energy density $\rho_\text{end}$.
Utilizing the relation $N_*\simeq q/(4\epsilon_{V*})$,  the formula for the number of e-folds during the reheating period (\ref{Nreh1}), $N_\text{reh}$, is recast into 
\begin{equation} \label{Nrehgr}
 \, N_\text{reh} (n_s,q, \bar{w}_\text{reh})\, \simeq \,  \frac{4}{1-3\bar{w}_\text{reh}} \left[ 57.5 - N_*(n_s) + \frac14 \ln \gamma  +  \frac{1-q/2}{4} \ln \frac{q}{4N_*(n_s)}\right]\,,   
\end{equation}
where
\begin{equation}
N_*(n_s)=\frac{q(1+n_s)+4}{4(1-n_s)}\,
\end{equation}
and a constant value for the effective equation of state, $\bar{w}_\text{reh}$, is assumed.
Accordingly, the reheating temperature (\ref{Treh1}) is written in terms of the model dependent parameters $q$, $\bar{w}_\text{reh}$ and the observable quantity $n_s$, 
\begin{equation} \label{Tgr}
T_\text{reh}(n_s, q, \bar{w}_\text{reh})= \left(\frac{1}{\gamma}\right)^{1/4} \lambda^{1/4}_q \left(\frac{q}{\sqrt{2}}\right)^{q/4} \left(\frac{30}{\pi^2 g_*}\right)^{1/4}  M_\text{Pl} \, e^{-\frac34(1+\bar{w}_\text{reh}) N_\text{reh}(n_s)}.
\end{equation}
We assume no violation of the null-energy condition, hence the expression in front of the exponential represents the maximum reheating temperature, $T_\text{max}=T_\text{max}(q, \lambda_q)$. For $N_\text{reh}>0$ the reheating temperature decreases exponentially with the $N_\text{reh}$.


\section{Inflaton with non-minimal derivative coupling}
We consider a theory of a scalar field $\phi$, which we identify with the inflaton, with non-minimal derivative coupling to gravity (NMDC).
The background dynamics of the scalar field $\phi$ are described by the Lagrangian
\begin{equation} \label{NMDC}
{\cal L}=\sqrt{-g} \left[\frac12 M^2_P R -\frac12\left(g^{\mu\nu} -\frac{G^{\mu\nu}}{\tilde{M}^2} \right)\partial_{\mu}\phi\partial_{\nu}\phi - V_\text{}(\phi)\right]\,.
\end{equation}
In a homogeneous FLRW background dominated by the inflaton $\phi$ the energy density and the pressure are respectively
\begin{equation} \label{rho}
\rho_\phi =\frac{\dot{\phi}^2}{2}\left(1+9\frac{H^2}{\tilde{M}^{2}}\right)+V_\text{}(\phi)
\end{equation}
and
\begin{equation} \label{p}
p_\phi =\frac{\dot{\phi}^2}{2}\left(1-3\frac{H^2}{\tilde{M}^{2}}\right) - V_\text{}(\phi) -\frac{1}{\tilde{M}^2} \frac{d(H\dot{\phi}^2)}{dt}.
\end{equation}
The Friedmann equation reads
\begin{equation} \label{fnmdc}
H^2=\frac{\rho_\phi}{3M^2_P}\,,
\end{equation}
and the equation of motion (EOM) for the NMDC $\phi$ field is
\begin{equation} \label{eos1}
\ddot{\phi} \left(1+3\frac{H^2}{\tilde{M}^2} \right) +3H\dot{\phi} \left(1+3\frac{H^2}{\tilde{M^2}} + \frac{2\dot{H}}{\tilde{M}^2}\right) + V'_\text{}(\phi)=0\,.
\end{equation}
The above equations yield the modified expressions for the slow-roll parameters \cite{Germani:2011ua}
\begin{equation} \label{gefslow}
\epsilon=\frac{\epsilon_V}{1+3H^2\tilde{M}^{-2}}\,,\quad\quad \eta=\frac{\eta_V}{1+3H^2\tilde{M}^{-2}}\,,
\end{equation}
where $\epsilon_V=(M^2_\text{Pl}/2)(V'/V)^2$ and $\eta_V=M^2_\text{Pl}(V''/V)$.

\begin{figure}
\textbf{ \quad\quad\quad $ V(\phi)=\frac12m^2_\phi \phi^2$ \quad\quad \quad\quad \quad\quad\quad \quad \quad \quad\quad \quad\quad  $V(\phi)=\lambda_\phi \phi^4$ }
\centering
\begin{tabular}{cc || cc}
{(a)} \includegraphics [scale=.62, angle=0]{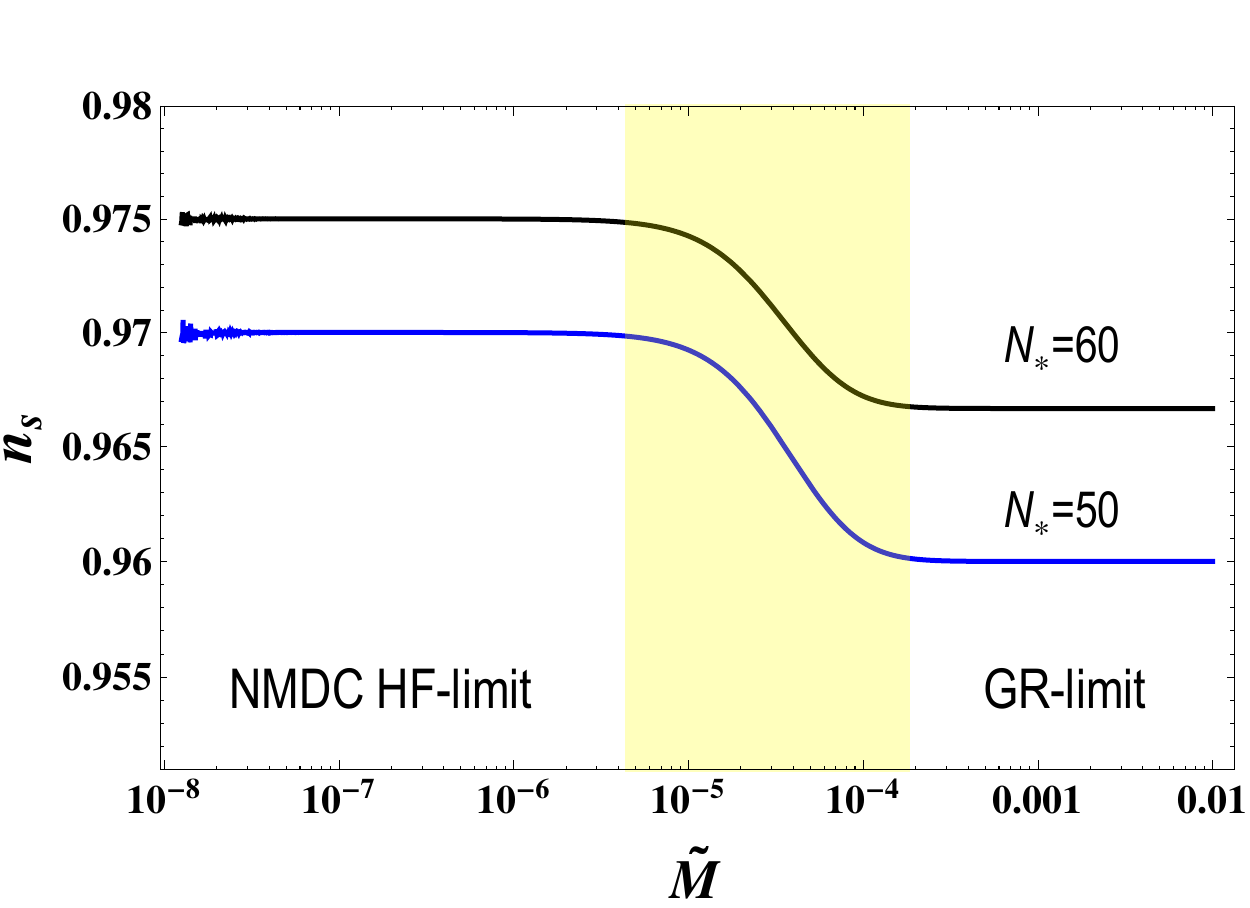} \quad
{(b)} \includegraphics [scale=.62, angle=0]{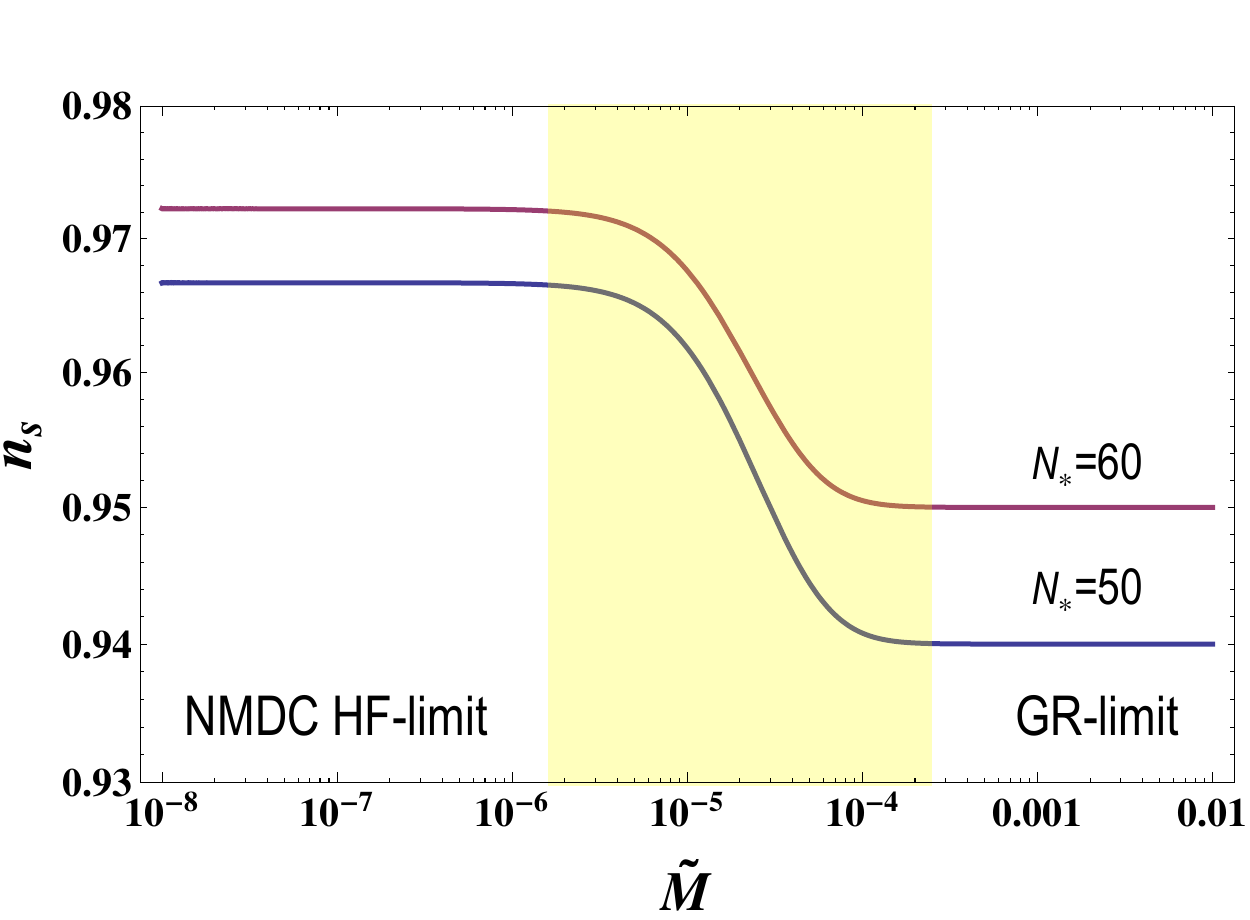}  \\
{(c)} \includegraphics [scale=.62, angle=0]{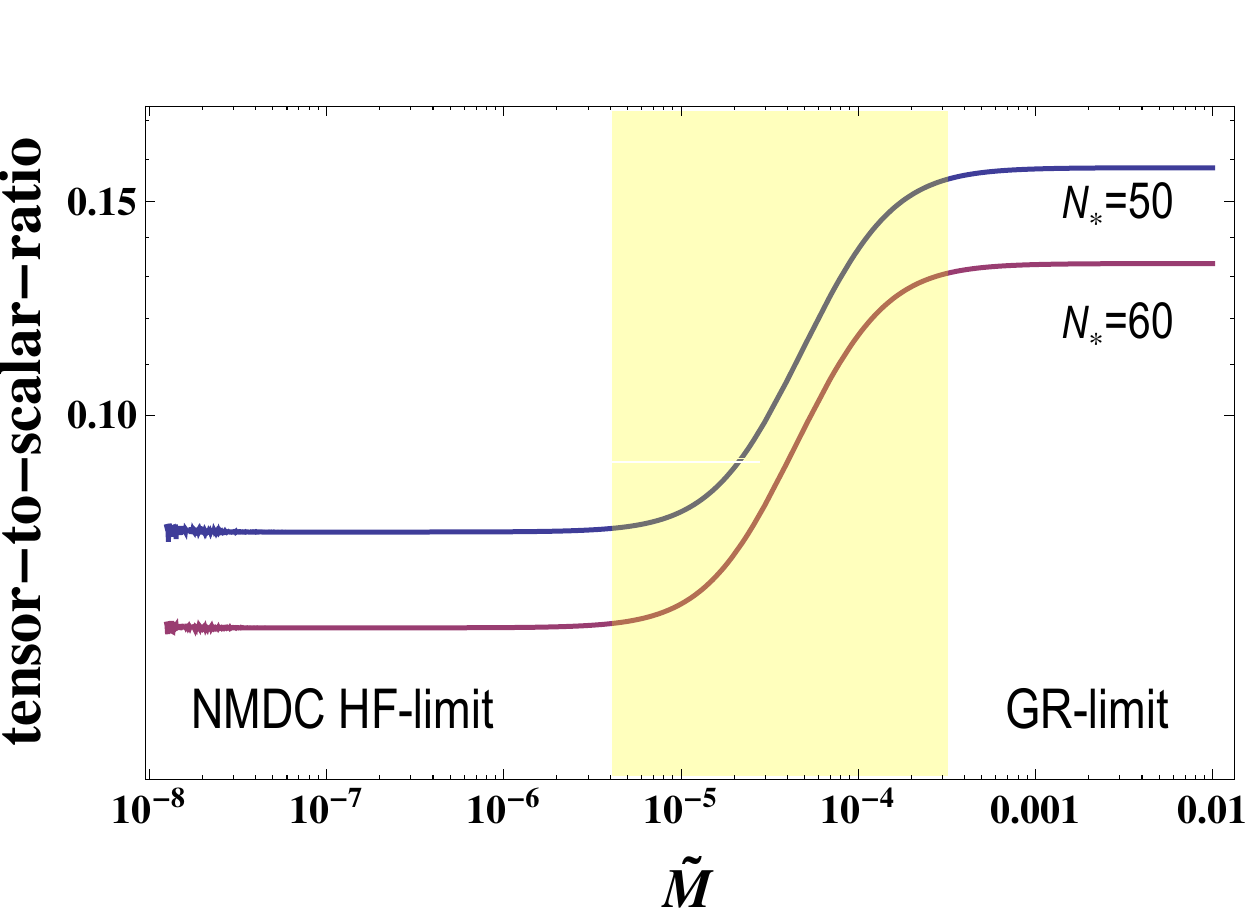} \quad
{(d)} \includegraphics [scale=.62, angle=0]{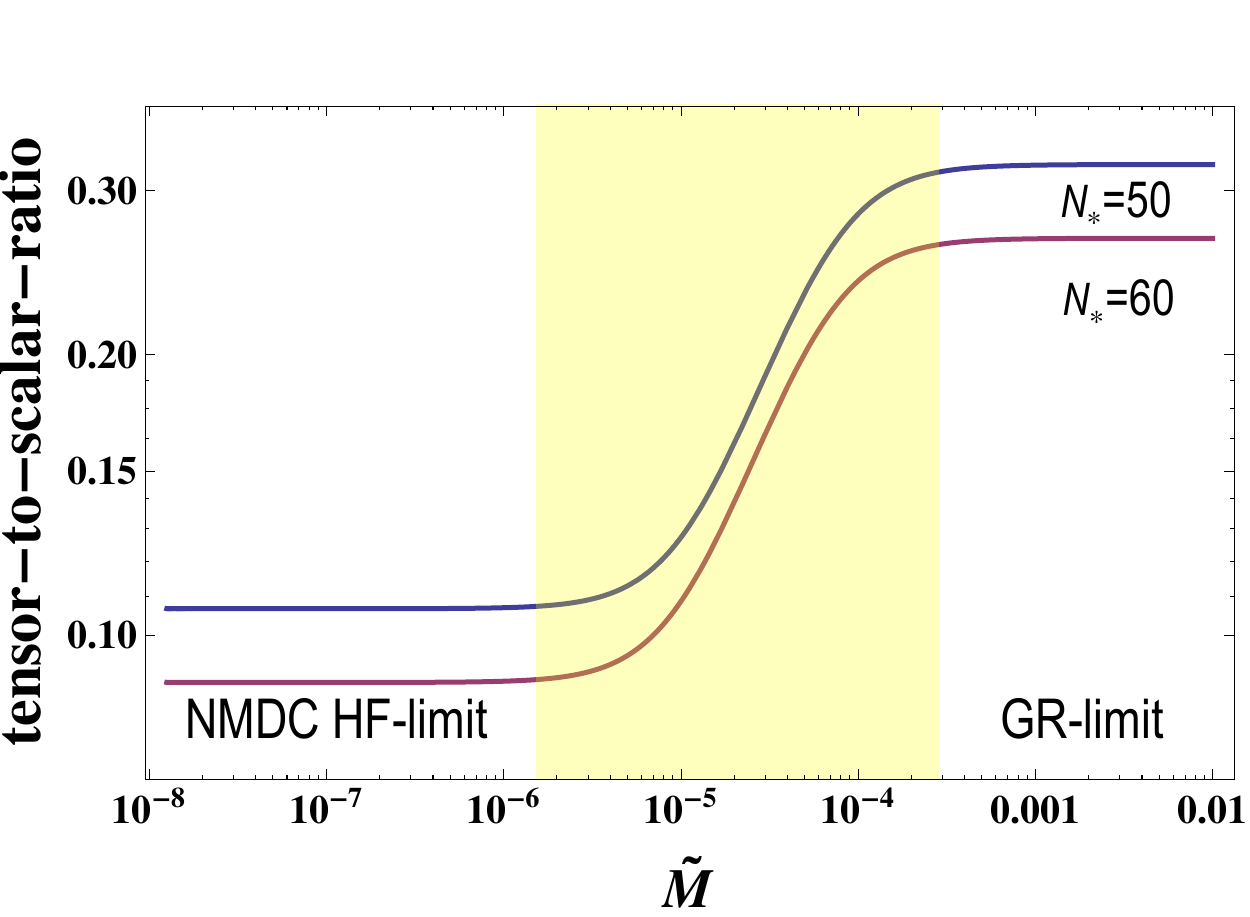}  \\
\end{tabular}
\caption{\small{ The change of the values for the CMB observables $n_s$ and tensor-to-scalar-ratio $r$ with respect to the $\tilde{M}$ value for the potentials $V=m^2_\phi \phi^2/2$ (left panels) and $\lambda_\phi \phi^4$ (right panels) and inflaton with NMDC. The e-folds number $N_*=50$ is shown in blue curve and $N_*=60$ in black. The plots clearly show three distinct regions according to the $\tilde{M}$ value where the NMDC is dominant (NMDC HF-limit), subdominant (GR-limit), and the intermediate region (light-yellow region). Mass dimensions are in Planck units.
}}
\end{figure}

The power spectrum of the primordial scalar perturbations written in terms of the slow roll parameters is modified, see Ref \cite{Tsujikawa:2012mk},
\begin{equation}  \label{modif}
\left. 1-n_s \equiv \frac{d \ln {\cal P}_\zeta}{d \ln k}\right|_{c_sk=aH}  \simeq  \frac{1}{1+3H^2\tilde{M}^{-2}} \left[2\epsilon_V \left(4-\frac{1}{1+3H^2\tilde{M}^{-2}} \right) -2\eta_V \right]\,.
\end{equation}
When the new  scale $\tilde{M}$ is much smaller than the Hubble  scale, i.e. $H\tilde{M}^{-1} \gg 1$, we have the so called {\it high friction limit} (HF-limit) or {\it gravitational enhanced friction} \cite{Germani:2011ua}. In the HF limit the spectral tilt of the scalar power spectrum gets simplified \cite{Germani:2010ux, Germani:2011ua},
\begin{equation} \label{ns}
 1-n_s  \simeq 8 {\epsilon} - 2{\eta}\,,
\end{equation}
albeit, the ${\cal P_\zeta}$ amplitude and in first order the consistency relation, $r=16 \epsilon$, remain unchanged. The high friction limit, $\tilde{M} \ll H_*$ is rather interesting because all the attractive features of the NMDC, such as the UV insensitivity to higher dimensional operators, even in the absence of symmetries \cite{Dalianis:2015aba},
and the good fit to the data, get evident.

For the large field models where the characteristic energy scale for inflation is $H_\text{inf} \sim 10^{-5} M_\text{Pl}$ we can distinguish three cases according to the $\tilde{M}$ value (see Fig. 2):
\begin{description}
	\item [i) ]$ 10^{-3} M_\text{Pl} \lesssim \tilde{M}$  (GR-limit)\\
	The NMDC plays essentially no role during the observed inflationary period. The $r(n_s)$ predictions, shown in Fig. 3, are given by the upper straight line in the $(n_s, r)$ plane.
	\item  [ii)] $10^{-6} M_\text{Pl} < \tilde{M} \lesssim 10^{-3} M_\text{Pl}$ (intermediate region)\\
	The NMDC modifies the inflationary dynamics and predictions but becomes negligible during the reheating period. The $r(n_s)$ predictions, lie in the yellow area of the  $(n_s, r)$ plane in Fig. 3.
	\item  [iii)] $10^{-\beta} M_\text{Pl} < \tilde{M} \lesssim 10^{-6} M_\text{Pl}$ (High Friction-limit)\\
	The NMDC modifies both the inflationary and reheating dynamics. The limit values for the power $\beta$, estimated by the CMB normalization, see Eq. (\ref{lp}), depend on the model. It is roughly $\beta=14, 9, 8$ for the linear, quadratic and quartic potential respectively. The $r(n_s)$ predictions are given by the lower straight line in the $(n_s, r)$ plane in Fig. 3.
\end{description}

A remarkable observation is that in the high friction limit, $\tilde{M} \ll H_\text{inf}$, and during the slow-roll phase the evolution of the field $\phi$ with potential $V(\phi)$  evolution resembles the evolution of a minimally coupled field $\varphi$  with potential $V_\text{GR}(\varphi)$. For the monomial potentials
\begin{equation} \label{pot1}
V(\phi) = \lambda_p\, M^4_\text{Pl} \left(\frac{\phi}{M_\text{Pl}} \right)^p
\end{equation}
there is the correspondence 
\begin{equation} \label{corre}
\left. \left. p\,\right|_\text{NMDC} \,\longleftrightarrow \,\frac{2p}{p+2}\right|_\text{GR}\,\equiv q\,,
\end{equation}
see Table 1.
The expressions for the spectral tilt $n_s(N_*)$ and the tensor-to-scalar ratio $r(N_*)$ for the NMDC models  \cite{Tsujikawa:2012mk} can be directly obtained by transforming the well known GR expressions  (\ref{A}) according to the relation (\ref{corre}),
\begin{equation} \label{A-nmdc}
1-n_s \simeq \frac{4(p+1)}{2(p+2)N_*+p}\,, \quad\quad r \simeq \frac{16\, p}{2(p+2)N_* + p} \,.
\end{equation}
Standard inflationary models such as the ($\phi^4, \varphi^{4/3}$), ($\phi^2, \varphi$) and ($\phi, \varphi^{2/3}$), as well as the ($e^\phi, \varphi^{2}$), can be seen as dual models yielding identical predictions in terms of the $r=r(n_s)$ relation.
We will show that this theoretical degeneracy can break when the cosmic evolution during the reheating stage is taken into account.

\begin{figure}
\centering
\begin{tabular}{cc || cc}
{(a)} \includegraphics [scale=.38, angle=0]{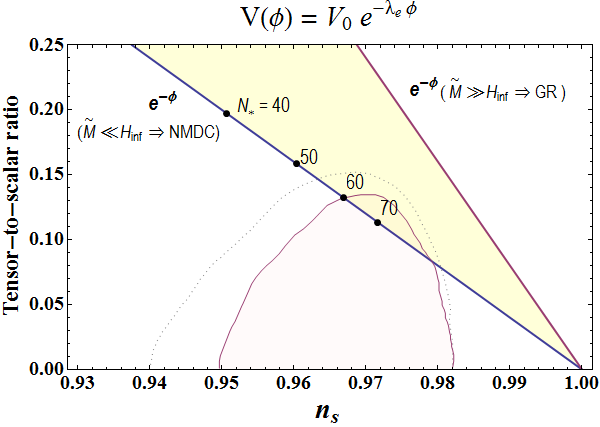} \quad
{(b)} \includegraphics [scale=.38, angle=0]{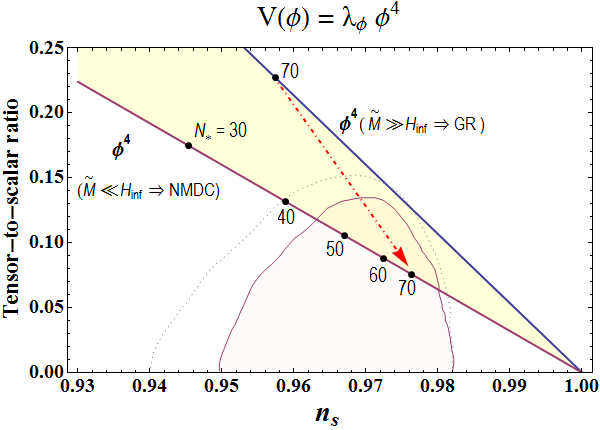}  \\
{(c)} \includegraphics [scale=.38, angle=0]{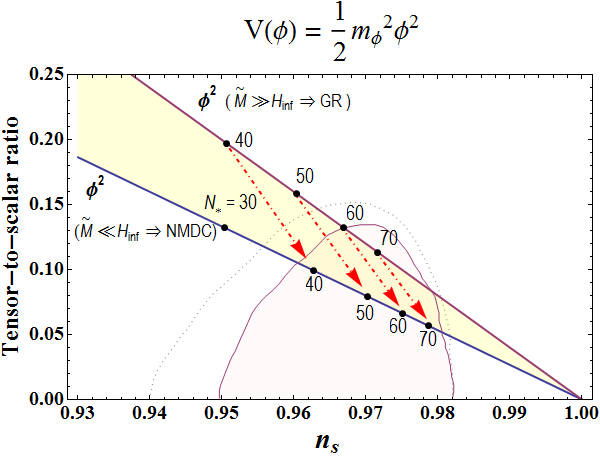} \quad
{(d)} \includegraphics [scale=.38, angle=0]{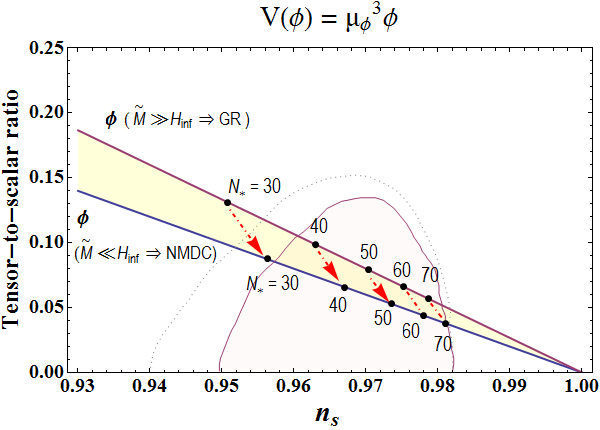}  \\
\end{tabular}
\caption{\small{ The marginalized joint $95\%$ CL regions for the  $n_s$ and the $r_{0.002}$ from {\itshape Planck} $2013$ (dotted) and from {\itshape Planck} TT$\_\text{lowP}$ $2015$.
The panels show the theoretical predictions for the exponential, quartic, quadratic and linear potential in the presence of NMDC. The upper straight line is the prediction for $r=r(n_s)$ with the NMDC effect negligible, i.e. $\tilde{M}\gg H_\text{inf}$; in this case each model effectively reduces to a canonical scalar field with Einstein gravity. The lower straight line is the prediction for $r=r(n_s)$ with the NMDC effect dominant, i.e. $\tilde{M}\ll H_\text{inf}$ ; in this case the predictions are much different than GR, and interestingly enough coincide with the predictions of a canonical scalar field with Einstein gravity but different potential.
The lightyellow region between the two theoretical straight lines captures the theoretical prediction $r=r(n_s)$ with intermediate $\tilde{M}$ values, $\tilde{M} \sim H_\text{inf}$.
}}
\end{figure}

In Hordenski theories the post inflationary  reheating stage, where the inflaton field dynamics dominate the universe evolution, may be significantly different from the minimal theories. The inflaton field velocity oscillates very fast and its value has to be bounded, $3\dot{\phi}^2 \leq 2 \tilde{M}^2 M^2_\text{Pl}$, for a positive definite potential, see Fig. 4.  Analysis of the inflaton oscillations in the class of models with NMDC has been also performed in a number of works \cite{Sadjadi:2012zp, Ghalee:2013ada, Sadjadi:2013na, Jinno:2013fka, Ema:2015oaa, Myung:2016twf, Ema:2016hlw}. Here we will follow the results of Refs. \cite{Jinno:2013fka, Ema:2015oaa}. There it was found that the Hubble parameter rapidly oscillates if the non-minimal kinetic term of the inflaton takes a dominant role. It was also mentioned that the sound speed squared of the scalar perturbation rapidly oscillates between positive and negative values and an instability for the shortest wavelength mode of the scalar perturbations may occur. 

On the other hand the super-horizon wavelength modes can be safely described using the perturbation theory for particular inflationary potentials.
It has been shown, in Ref. \cite{Germani:2015plv}, that for the $\phi^4$ potential the linear regime is well possible to remain valid in theories with NMDC. The linear regime for the superhorizon modes is essential for our discussion; if it was otherwise the CMB observables could not be linked with the inflationary universe predictions in a calculable manner.

\section{Inflationary observables with Non-Minimal Derivative Coupling}

It is rather interesting that inflationary models with {\itshape monomial potentials}, such as the $\phi^4$ and the $\phi^2$, or the {\itshape exponential potential} can successfully fit the observational data when the higher derivative coupling (\ref{NMDC}) is in action. We will investigate the inflationary observables for the potentials
\begin{equation}  \label{potDC}
V(\phi) = \lambda_p\, M^4_\text{Pl} \left(\frac{\phi}{M_\text{Pl}} \right)^p\,,\quad\quad\quad V(\phi)=V_0 e^{-\lambda_e\phi/M_\text{Pl}} \,.
\end{equation}
We consider the {\itshape high friction} limit \cite{Germani:2011ua} where $H_*\gg \tilde{M}$ and the Eq. (\ref{gefslow}) approximates into $\epsilon\simeq \epsilon_V/(3H^2\tilde{M}^{-2})$, $\eta \simeq \eta_V/(3H^2\tilde{M}^{-2})$.  For the potentials (\ref{potDC}) we take respectively the expression for the first slow-roll parameter in terms of the inflaton field value, $\epsilon(\phi)= ({p^2}/{2\lambda_p}) ({\tilde{M}^{2} M^p_\text{Pl}}/{\phi^{p+2}})$ and $\epsilon(\phi)=(\lambda^2_e \tilde{M}^2/2V_0) e^{\lambda_e \phi/M_P}$.
The $\phi_\text{end}$ is approximately obtained from setting the first slow-roll parameter, $\epsilon=\epsilon_V/(3H^2_\text{end}\tilde{M}^{-2})$, equal to one. We write the energy density at the end of inflation in terms of the potential as $3H^2_\text{end} M^2_\text{Pl} =\rho_\text{end}\equiv \gamma^{-1} V_\text{end}$ and we get
\begin{equation}
\phi^{p+2}_\text{end} = ({\gamma\,p^2}/{2\lambda_p}) \tilde{M}^2 M^p_\text{Pl}\,,
\end{equation}
\begin{equation}
\phi_\text{end}=(M_\text{Pl}/\lambda_e)\ln[2V_0/(\gamma\, \lambda^2_e\tilde{M}^2 M^2_\text{Pl})]
\end{equation}
for the two classes of potentials (\ref{potDC}) respectively. We find that the first slow-roll parameter becomes of order unity when $V_\text{end}/\rho_\text{end}\equiv \gamma\sim 2/3$ and it is generally $\gamma<2/3$ when the $\phi^p$ inflation terminates, see Fig. 9. 
Actually, the reheating e-folds number is only logarithmically sensitive to the $\gamma$-factor value, see Eq. (\ref{Ndc}).

\begin{figure}
\centering
\begin{tabular}{cc}
{(a)} \includegraphics [scale=.5, angle=0]{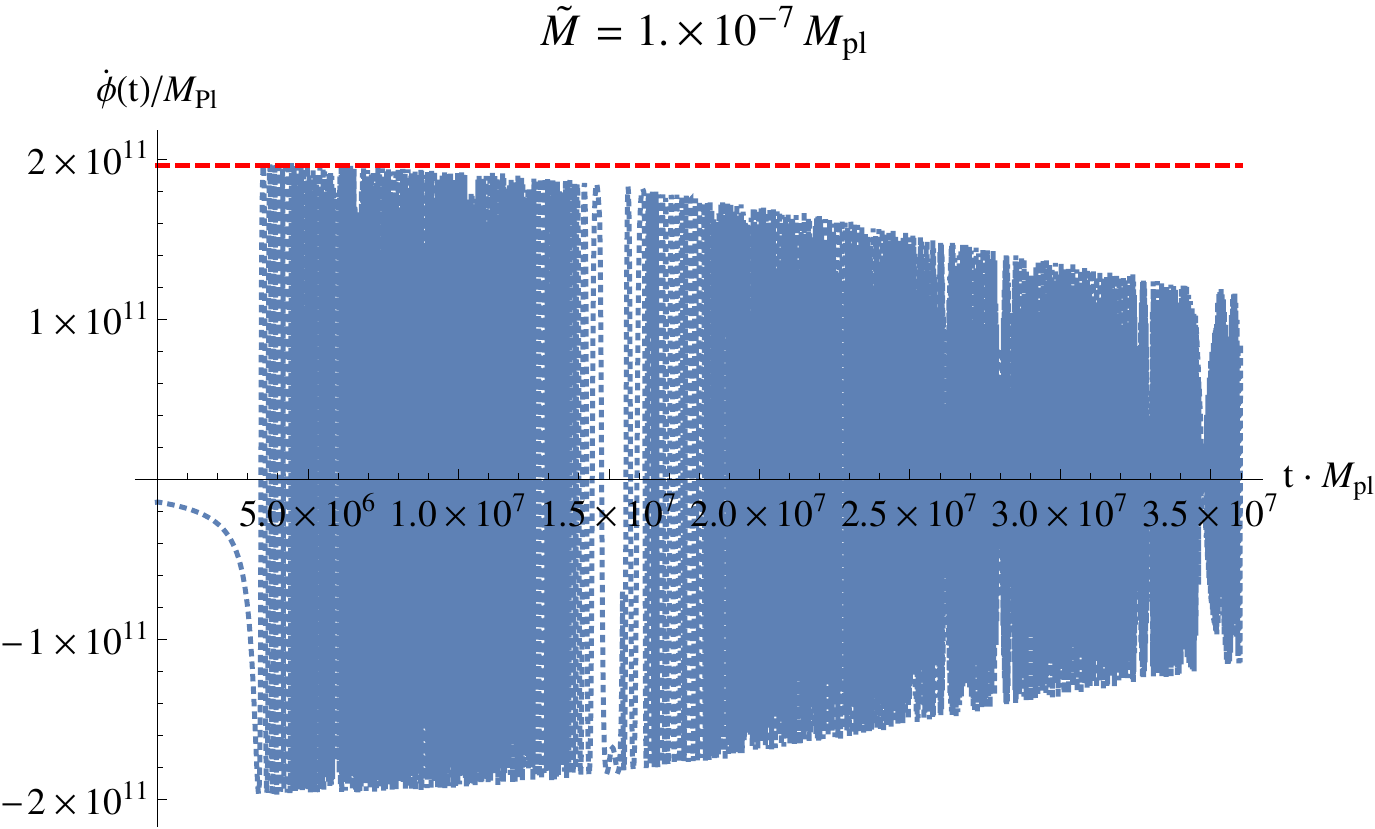} \quad
{(b)} \includegraphics [scale=.5, angle=0]{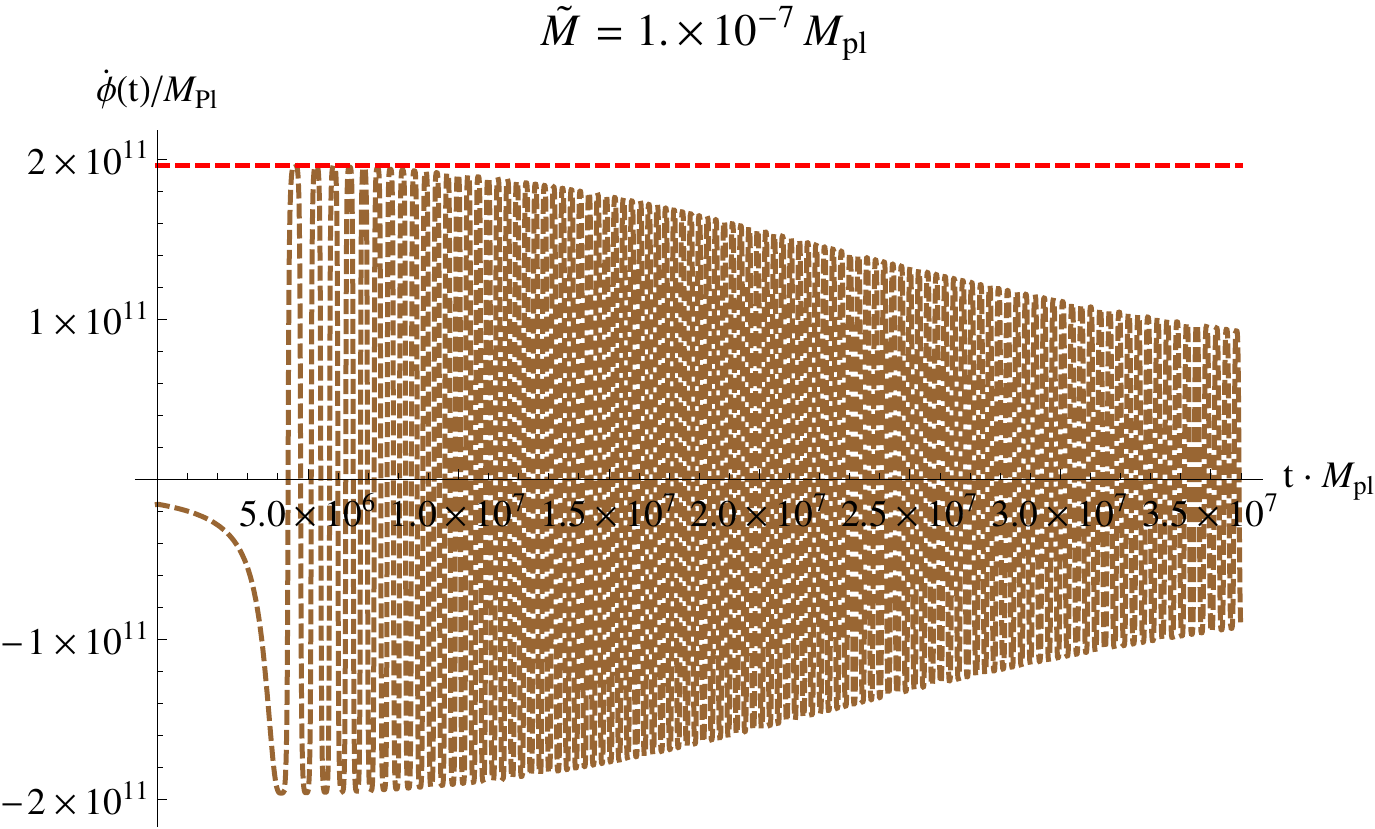}  \\
\end{tabular}
\caption{\small{ The rapid oscillations of the inflaton field velocity for quadratic (left panel) and quartic (right panel) potentials when the NMDC dominates, $\tilde{M}=10^{-7} M_\text{Pl}$. The red-dashed line shows the maximum allowed value for the $\dot{\phi}$.
}}
\end{figure}

The number of e-folds from the moment the mode $k_*$ exited the horizon until the end of inflation is
\begin{equation} \nonumber
N_*\equiv\int^{t_{end}}_{t_*} Hdt =\frac{1}{M^2_\text{Pl}} \int^{\phi_*}_{\phi_{end}}(1+3H^2\tilde{M}^{-2}) \frac{V}{V'}d\phi,
\end{equation}
where the slow-roll approximation of the EOM (\ref{eos1}), $3H(1+3H^2\tilde{M}^{-2})\dot{\phi}=-V'(\phi)$, was used. In the {\itshape high friction} limit, $H_*\gg \tilde{M}$, the value of the inflaton, $N$-efolds before the end of inflation  is
\begin{equation} \label{phiDC}
\phi^{p+2}(N) = \frac{2p(p+2)N + \gamma \, p^2}{2\lambda_p} \tilde{M}^2 M^p_\text{Pl}\,,\quad\quad  \phi(N) =\frac{1}{\lambda_e} \ln \left(\frac{2 V_0}{\gamma \,\lambda^2_e \tilde{M}^2 N} \right)\,.
\end{equation}
for the two classes of potentials (\ref{potDC}) respectively.
We observe that for $\tilde{M} \rightarrow 0$ the inflationary trajectory shrinks well below Planck values, $\Delta \phi \ll M_\text{Pl}$, and sub-Planckian excursions of the (non-canonical) inflaton can adequately inflate the universe.

For the {\itshape monomial potentials} the slow-roll parameters $\epsilon$ and $\eta$ in the HF limit at the moment of horizon exit 
read
\begin{equation} \label{epet}
\epsilon_* \simeq \frac{\epsilon_{V*}}{3H^2_*\tilde{M}^{-2}} = \frac{p^2}{2\lambda_p} \frac{\tilde{M}^{2} M^p_\text{Pl}}{\phi^{p+2}_*} \simeq \frac{p}{2(p+2)N_*+ \gamma \, p}\,, \quad\quad \eta_* =\frac{2p-2}{p}\epsilon_*\,.
\end{equation}
Utilizing the above expressions, the inflationary observables can be obtained.  The spectral tilt of the scalar power spectrum and the tensor-to-scalar-ratio (\ref{ns})  read
\begin{equation} \label{ns2}
 1-n_s  \simeq 8 {\epsilon}_* - 2{\eta}_* =  \frac{4(p+1)}{2(p+2)N_*+ \gamma \, p}\,, \quad\quad r \simeq \frac{16\, p}{2(p+2)N_* + \gamma\, p} \,.
\end{equation}
The (slight) difference with the expression (\ref{A-nmdc}), derived using the slow-roll approximation, is the presence of the $\gamma$ factor due to the exit from the slow-roll phase. Interestingly enough, the inflationary observables for the {\itshape the exponential potential} are also obtained in the limit $p\rightarrow \infty$ \cite{Dalianis:2014nwa}.

\begin{figure}
\centering
\begin{tabular}{cc}
{(a)} \includegraphics [scale=.62, angle=0]{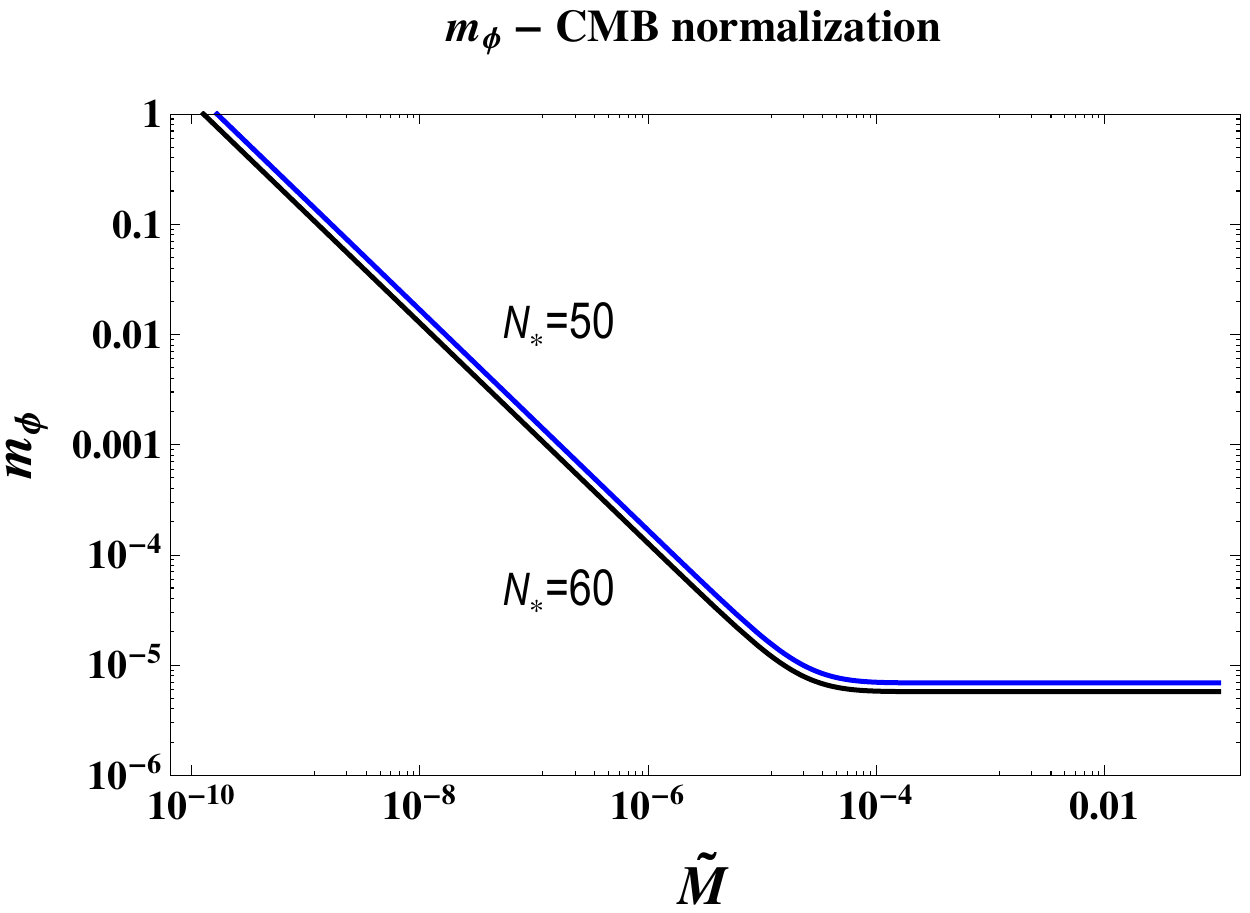} \quad
{(b)} \includegraphics [scale=.62, angle=0]{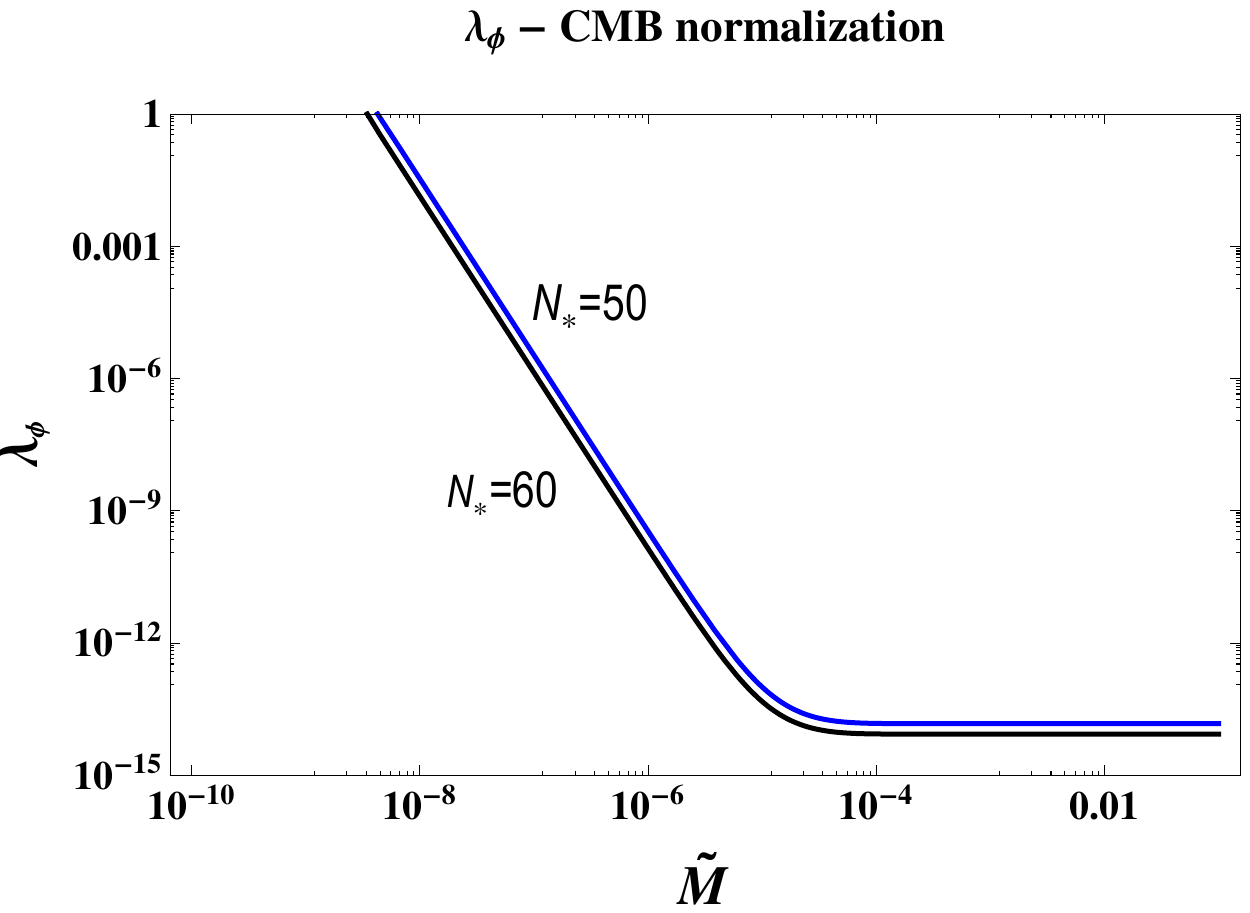}  \\
\end{tabular}
\caption{\small{ The CMB normalized values for the $m_\phi$ (left panel) and $\lambda_\phi$ (right panel), for $N_*=50$ in blue and $60$ in black, with respect to the $\tilde{M}$ value for the potentials $V=m^2_\phi \phi^2/2$ and $\lambda_\phi \phi^4$ and inflaton with NMDC. Remarkably large values for the $m_\phi$ and $\lambda_\phi$ become compatible with observations.  Note that the plots indicate a minimum value for the $\tilde{M}$ where the inflationary models are reliable and a maximum value for the $\tilde{M}$,  where the NMDC effects become negligible. Mass dimensions are in Planck units.
}}
\end{figure}

\subsection{CMB Normalization}

The power spectrum, measured at the scale $k_*$ by the CMB observational probes, reads in the NMDC case
\begin{equation} \label{Power'}
{\cal P}_\zeta =\frac{H^2_*}{8\pi^2 \epsilon_* c_s M^2_\text{Pl}} \simeq \frac{V_\text{}(\phi_*)}{24\pi^2 \epsilon_* c_s M^4_\text{Pl}}\,,
\end{equation}
where the second equality follows from the slow-roll approximation.
Plugging in the expressions for the potentials and utilizing the high friction result (\ref{phiDC}) we acquire the CMB normalized value for the (dimensionless) coefficients $\lambda_p(N_*, \tilde{M})$ and $\lambda_e(N_*, \tilde{M})$ in the NMDC scenario
\begin{equation} \label{lp}
\lambda_p(N_*, \tilde{M}) = \frac{ 24 \pi^2 c_s\, p} {2(p+2)N_*+p} \frac{M^p_\text{Pl}}{\left[\frac{2p(p+2)N_*+p^2}{2\lambda_p} \tilde{M}^2M^p_\text{Pl} \right]^{p/(p+2)}}\, {\cal P}_\zeta\,,
\end{equation}
\begin{equation} \label{le}
\lambda_e(N_*, \tilde{M}) = \sqrt{12}\,\pi\, \frac{M_\text{Pl}}{\tilde{M} N_*} \, {\cal P}^{1/2}_\zeta\,.
\end{equation}
The above expressions for the $\lambda_p$ and $\lambda_e$ coefficients correspond to an inflaton with NMDC, that is an inflaton with non-canonical kinetic term. After the inflaton field gets canonicalized these coefficients change accordingly. The $\lambda_p$ and $\lambda_e$ are the original potential parameters that appear in the NMDC Lagrangian (\ref{NMDC1}) and can be considered as the actual physical quantities only after the NMDC becomes negligible, something that happens during the reheating stage. 

\subsubsection{Standard examples}
For the {\itshape linear potentials}, $V_\text{}(\phi)=\mu^3_\phi  \phi$ it is $\lambda_1=\mu^3/M^3_\text{Pl}$. We obtain from Eq. (\ref{lp})
the expression for the parameter $\mu^3_\phi$
\begin{equation}
\mu^3_\phi(N_*, \tilde{M}) =0.52 \times 10^{-14}   \left(\frac{50}{N_*} \right)^2 \left(\frac{{\cal P}_\zeta}{2\times 10^{-9}} \right)^{3/2} \frac{M^4_\text{Pl}}{\tilde{M}}\,,
\end{equation}
where we took $c_s\sim 1$. An IR completion of the $V_\text{}(\phi)=\mu^3_\phi \phi$ model is of the form $V_\text{}(\phi)=\mu^3_\phi \left(\sqrt{\phi^2+\phi^2_c}-\phi_c \right)$, see e.g Ref. \cite{Adshead:2015pva}. 
Field values $\phi\ll \phi_c \ll M_\text{Pl}$ experience the low energy potential $V_\text{}(\phi)=(1/2)(\mu^3/\phi_c)  \phi^2$.  The low energy effective mass squared of the (non-canonical) inflaton is $m^3/\phi_c$ and its value can be closer to the Planck scale due to the NMDC. The effective mass of the canonicalized inflaton is suppressed by the factor $\tilde{M}/H$ for $H/\tilde{M} \gg 1$ and it is an increasing function with time due to the decrease of the $H(t)$ \cite{Ema:2015oaa}.

For the {\itshape quadratic potentials}, $V_\text{}(\phi)=m^2_\phi \phi^2/2$, it is $\lambda_2=m^2_\phi/(2M^2_\text{Pl})$. The Eq. (\ref{lp}) gives the mass $m_\phi$ with respect to the number of e-folds and the mass scale $\tilde{M}$
\begin{equation}
m_\phi(N_*, \tilde{M}) = 1.7 \times 10^{-10} \left(\frac{50}{N_*}\right)^{3/2} \left(\frac {{\cal P}_\zeta}{2\times 10^{-9}}\right)  \frac {M^2_\text{Pl}}{ \tilde{M}}\,,
\end{equation}
for $c_s\sim 1$. The high friction limit  $\tilde{M} \ll H_* \sim 10^{-5}M_\text{Pl}$ implies that the (non-canonical) inflaton mass is $m_\phi \gg 10^{-6} M_\text{Pl}$,
see Fig. 5, with the energy density of the inflaton remaining sub-Planckian. During inflation the canonical inflaton mass takes the standard value, about $6\times 10^{-6} M_\text{Pl}$. After inflaton the Hubble scale decreases and for $\tilde{M}<H$ the canonical effective mass, which can be defined as $m_\phi \times \tilde{M}/H$, is an increasing function with time implying that particles with mass larger than $6\times 10^{-6} M_\text{Pl}$ {\itshape can} be  kinematically produced via the perturbative decay of the inflaton. This fact  also makes high reheating temperatures feasible.

For the {\itshape quartic potentials}, $V_\text{}(\phi)=\lambda_\phi \phi^4$, it is $\lambda_4=\lambda_\phi$ and the scale $\tilde{M}$ is already constrained by the $(n_s, r)$ contour to be $\lambda_\phi M^2_\text{Pl}/\tilde{M}^2 >9.0 \times 10^{-5}$ \cite{Tsujikawa:2013ila,Tsujikawa:2012mk}. From the CMB normalization (\ref{lp}) we deduce that
\begin{equation} \label{eqlam}
\lambda_\phi(N_*, \tilde{M})\, \simeq \, 2.3 \times 10^{-32} \,  \left(\frac{50}{N_*}\right)^5 \left(\frac {{\cal P}_\zeta}{2\times 10^{-9}}\right) \left(\frac{M_\text{Pl}}{\tilde{M}}\right)^4\,,
\end{equation}
and self-coupling values $\lambda_\phi \gg 10^{-14}$ for the (non-canonical) inflaton are possible, see Fig. 5. Accordingly here, for $\tilde{M}<H$  the effective self-coupling for the canonical inflaton could be defined as $\lambda_\phi \times (\tilde{M}/H)^4$ which is an increasing function with time during the reheating period reaching the value $\lambda_\phi$ at the time the NMDC effects become negligible.

Finally, for the {\itshape exponential potentials} we obtain from (\ref{le}) the coefficient value $\lambda_e$ written in terms of $N_*$ and $\tilde{M}$
\begin{equation}
\lambda_e(N_*, \tilde{M})\simeq 10^{-5}\,\left(\frac{50}{N_*}\right) \left( \frac{{\cal P}_\zeta}{2\times 10^{-9}}\right) \frac{M_\text{Pl}}{\tilde{M}}\,.
\end{equation}

\subsection{The reheating stage}
The post inflationary evolution of the Horndeski theories is much different than the minimal case. In theories with NMDC the energy density and the pressure are given by the non-standard expressions (\ref{rho}) and (\ref{p}). The inflaton oscillates coherently and very fast about the minimum with the frequency $\omega_\text{eff} \sim (\tilde{M}/H) (V'/\phi)^{1/2}$ \cite{Ema:2015oaa}. The dynamics of the system $g_{\mu\nu}$ and $\phi$ are complicated. The expansion rate $H$ oscillates very fast and the $\rho_\phi$ is not a conserved quantity in an oscillation time scale. In the work of Ref. \cite{Jinno:2013fka, Ema:2015oaa} the quantity $J=H^{-1}\left[\left(1+6H^2/\tilde{M}^2 \right)\dot\phi^2/2+V\right]$ was employed and the averaged expansion law of the Universe was estimated to be
\begin{equation} \label{Ht}
\bar{H}\, \sim \,\frac{2p+2}{3p} \frac1t\,
\end{equation}
for {\itshape monomial} inflationary potentials $V(\phi)\propto \phi^p$. 
From the energy conservation equation and the definition of the averaged EoS during the reheating period (\ref{avEoS}) follows that the averaged energy density scales as $\bar{\rho}(a, \bar{w}_\text{reh})=\rho_\text{end}(a/a_\text{end})^{-3(1+\bar{w}_\text{reh})}$ and 
the averaged Hubble rate as
\begin{equation}
\bar{H}(t, \bar{w}_\text{reh})=\frac{1}{\frac{3}{2}(1+\bar{w}_\text{reh})(t-t_\text{end}) +H^{-1}_\text{end}}
\end{equation}
for $t_\text{end} < t< t_\text{reh}$.
  
\begin{figure}
\centering
\begin{tabular}{cc || cc || cc}
\includegraphics [scale=.58, angle=0]{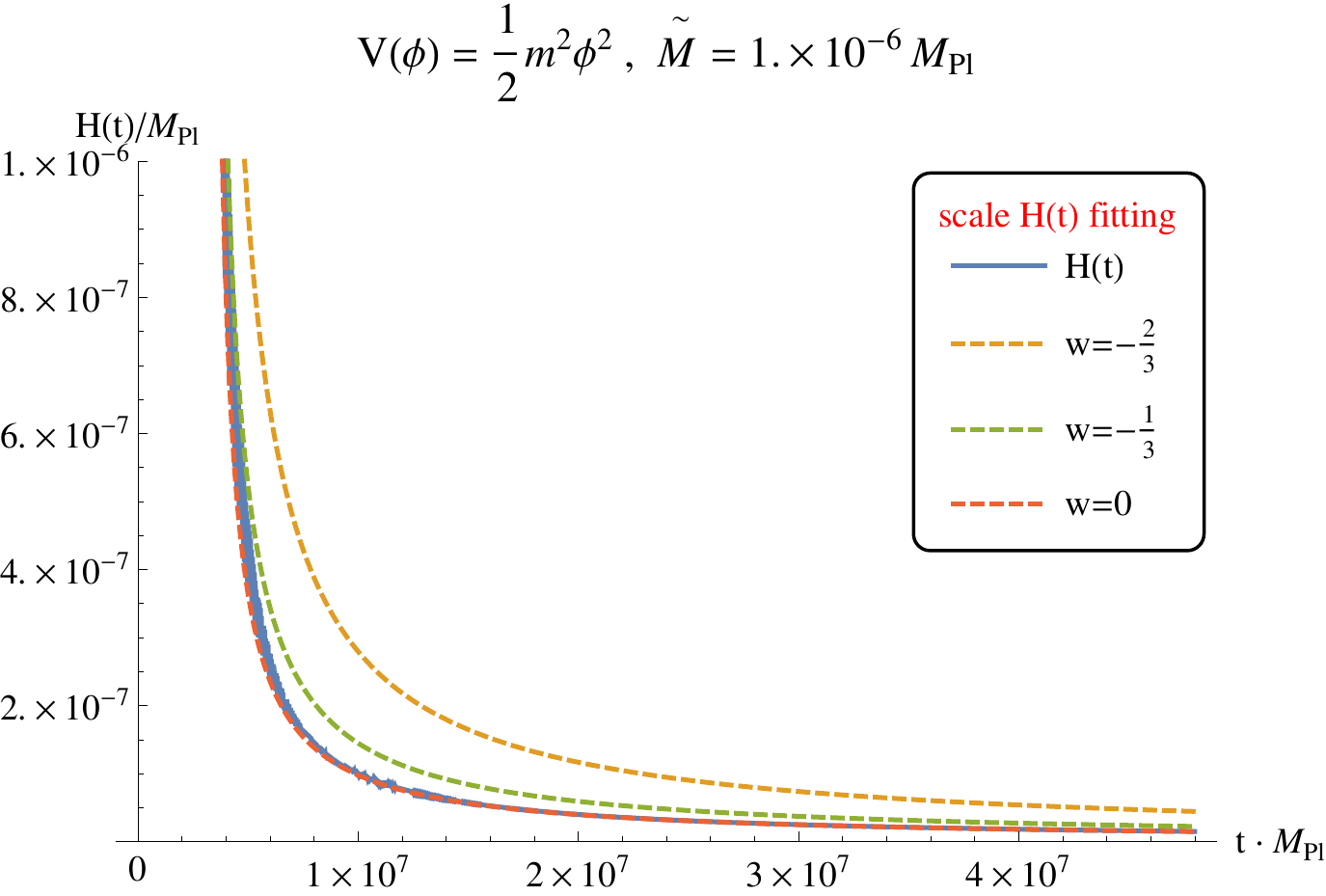} \quad
\includegraphics [scale=.58, angle=0]{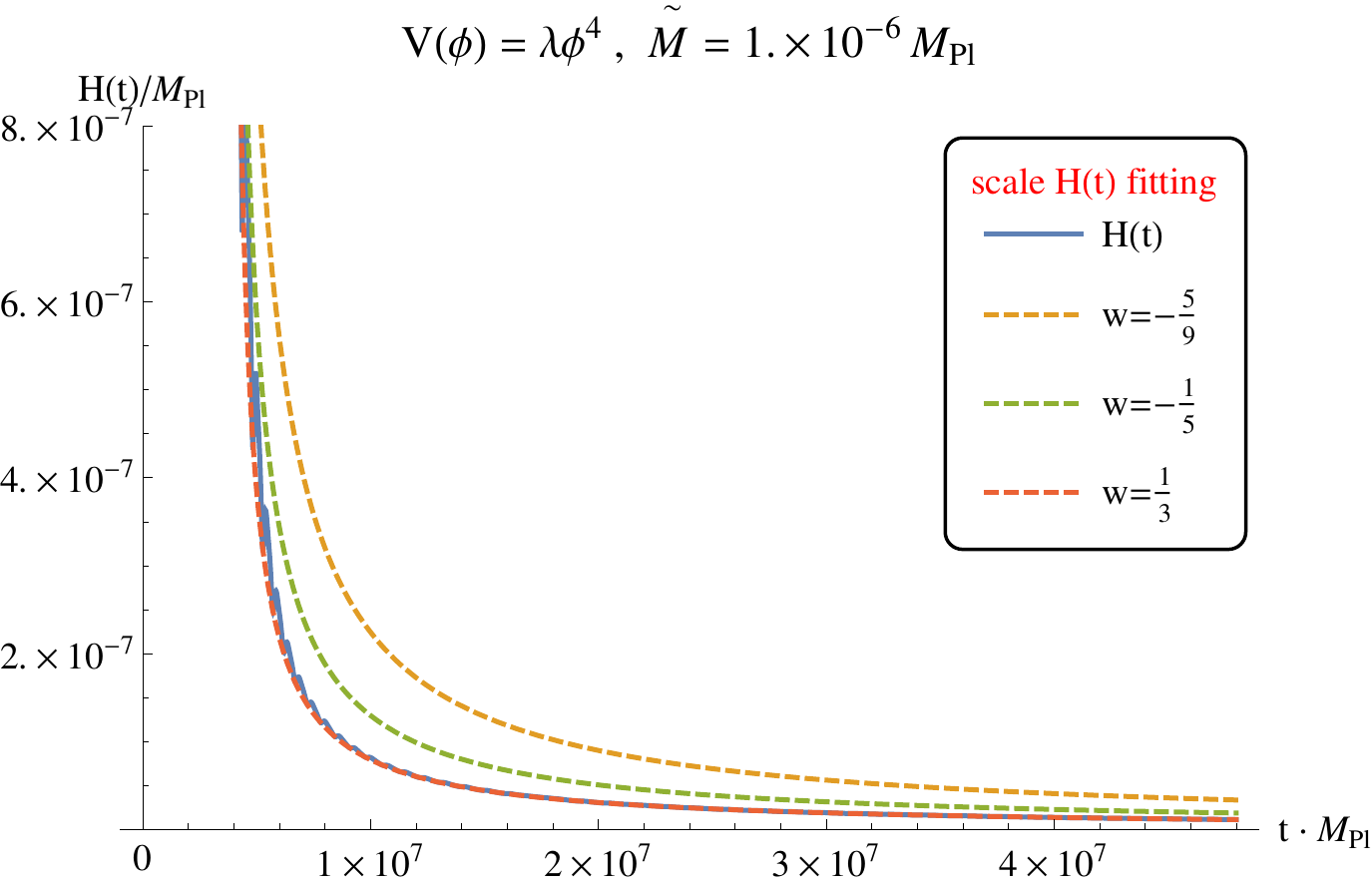} \\
\includegraphics [scale=.58, angle=0]{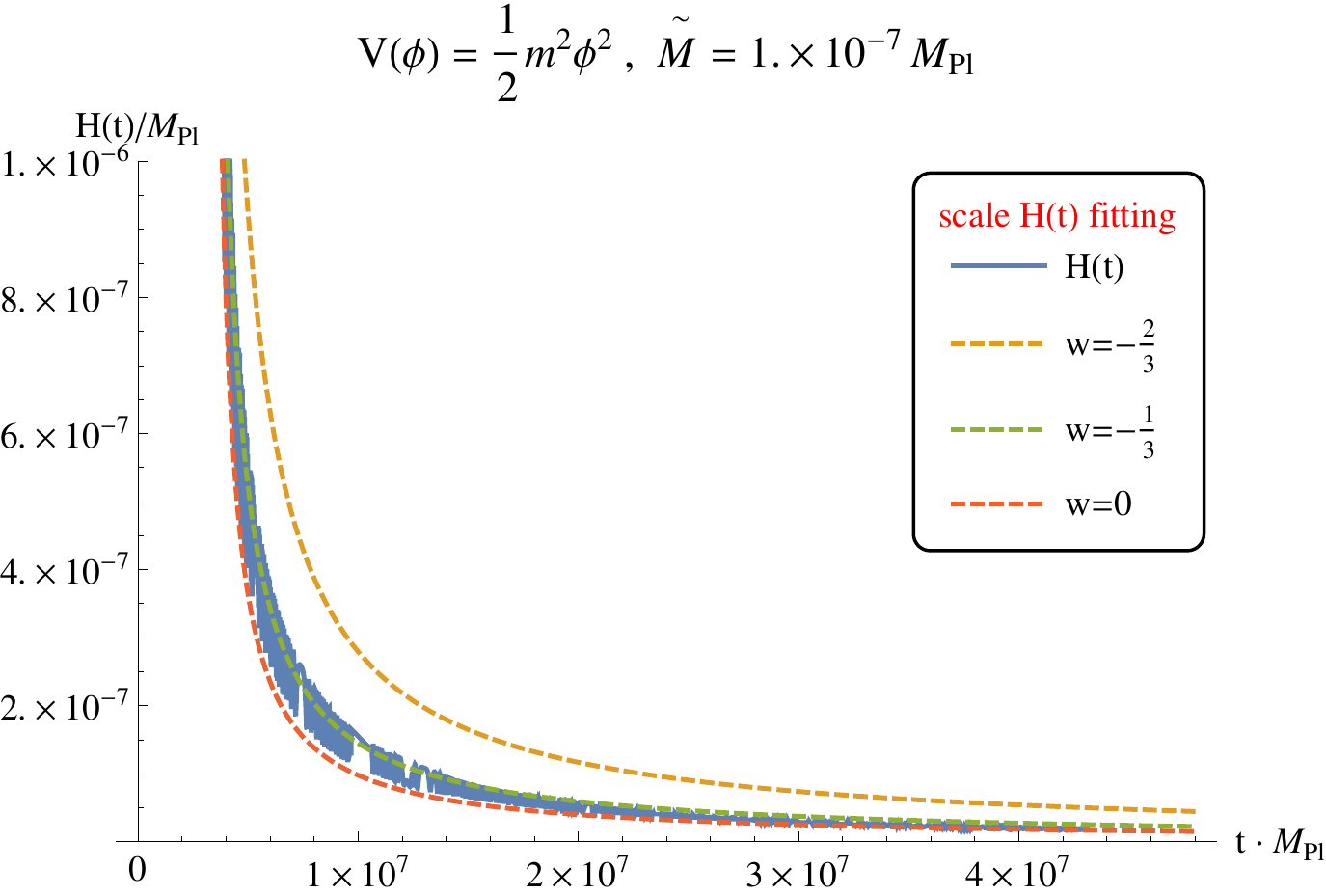} \quad
\includegraphics [scale=.58, angle=0]{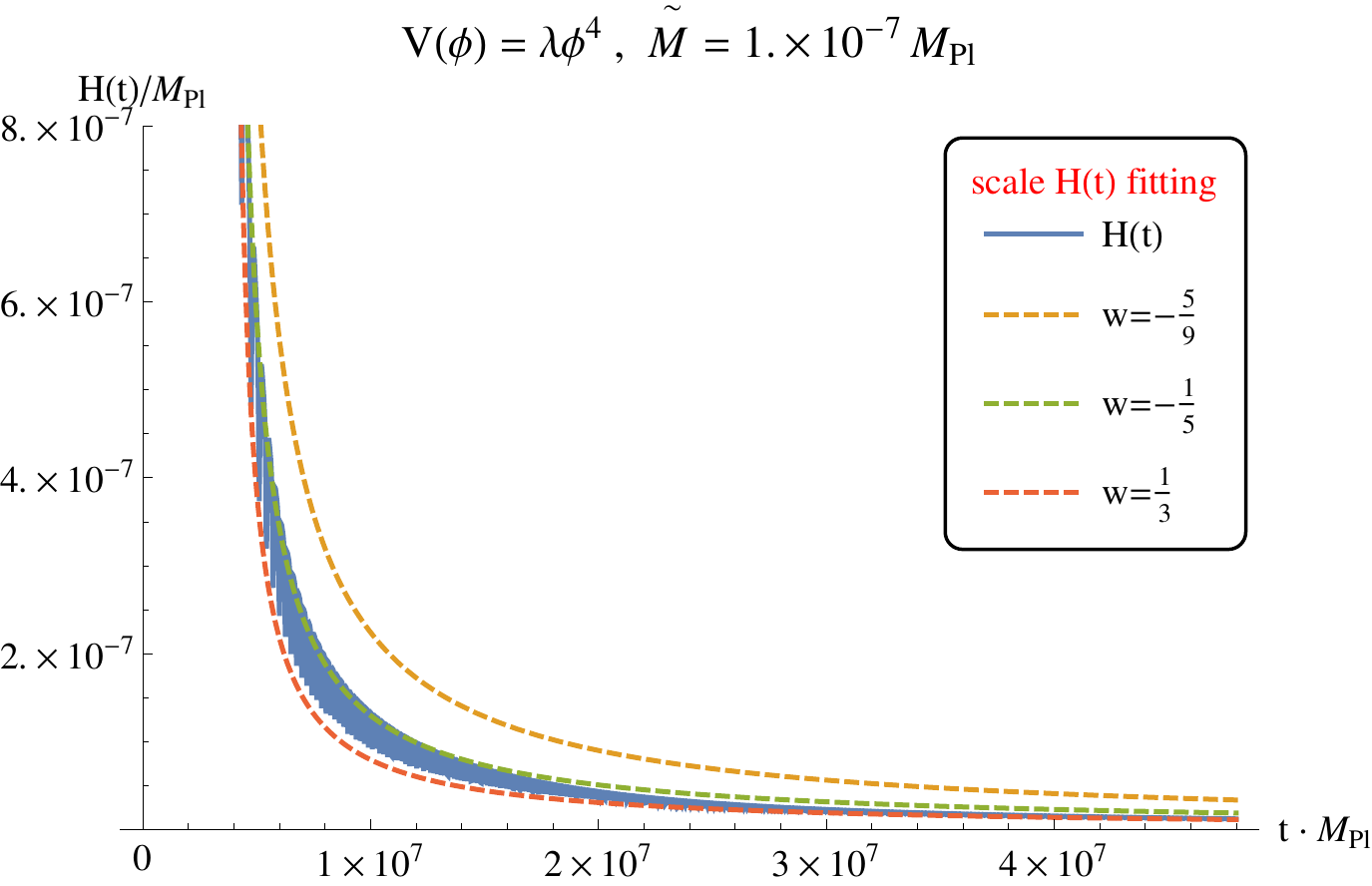} \\
\includegraphics [scale=.58, angle=0]{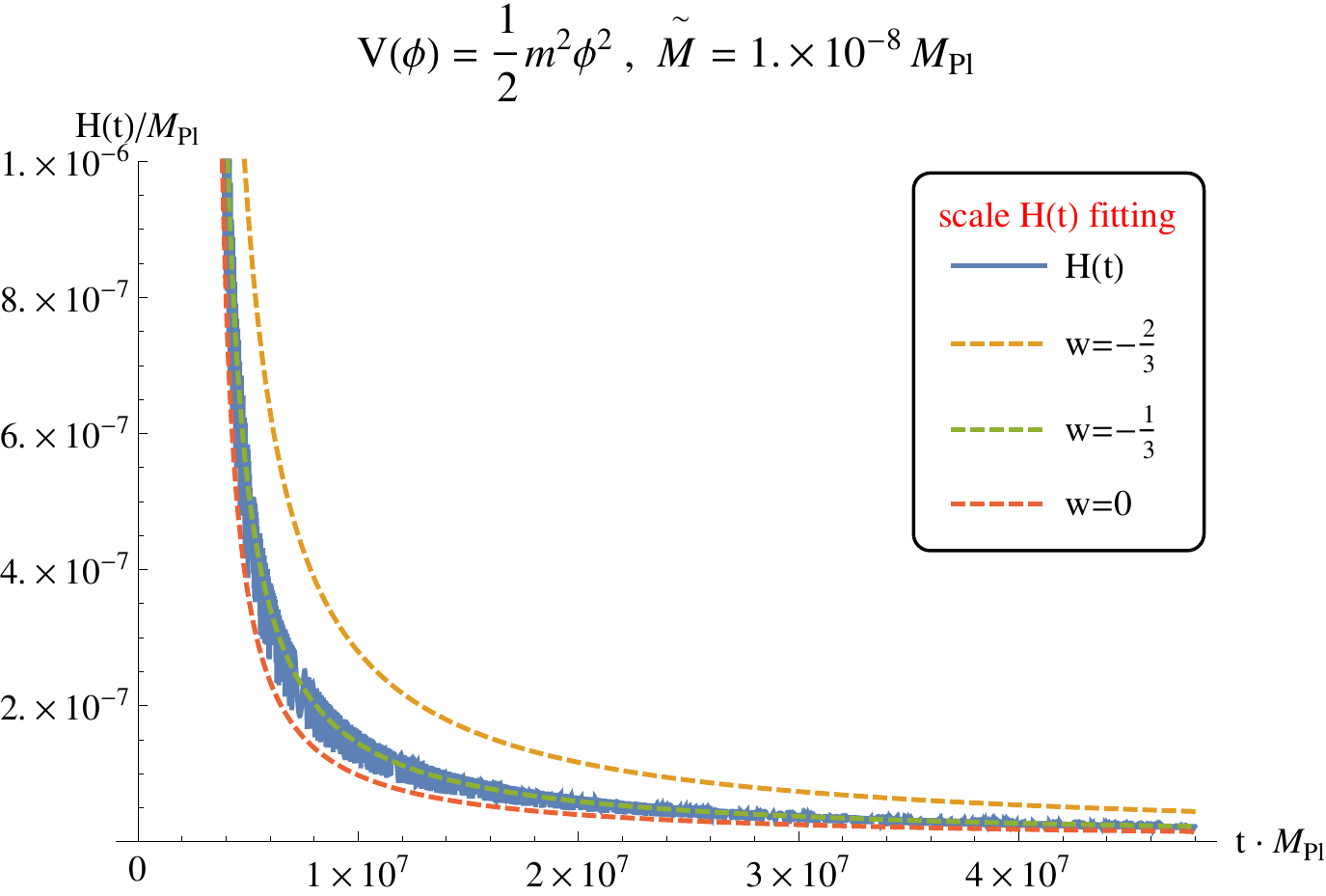} \quad
\includegraphics [scale=.58, angle=0]{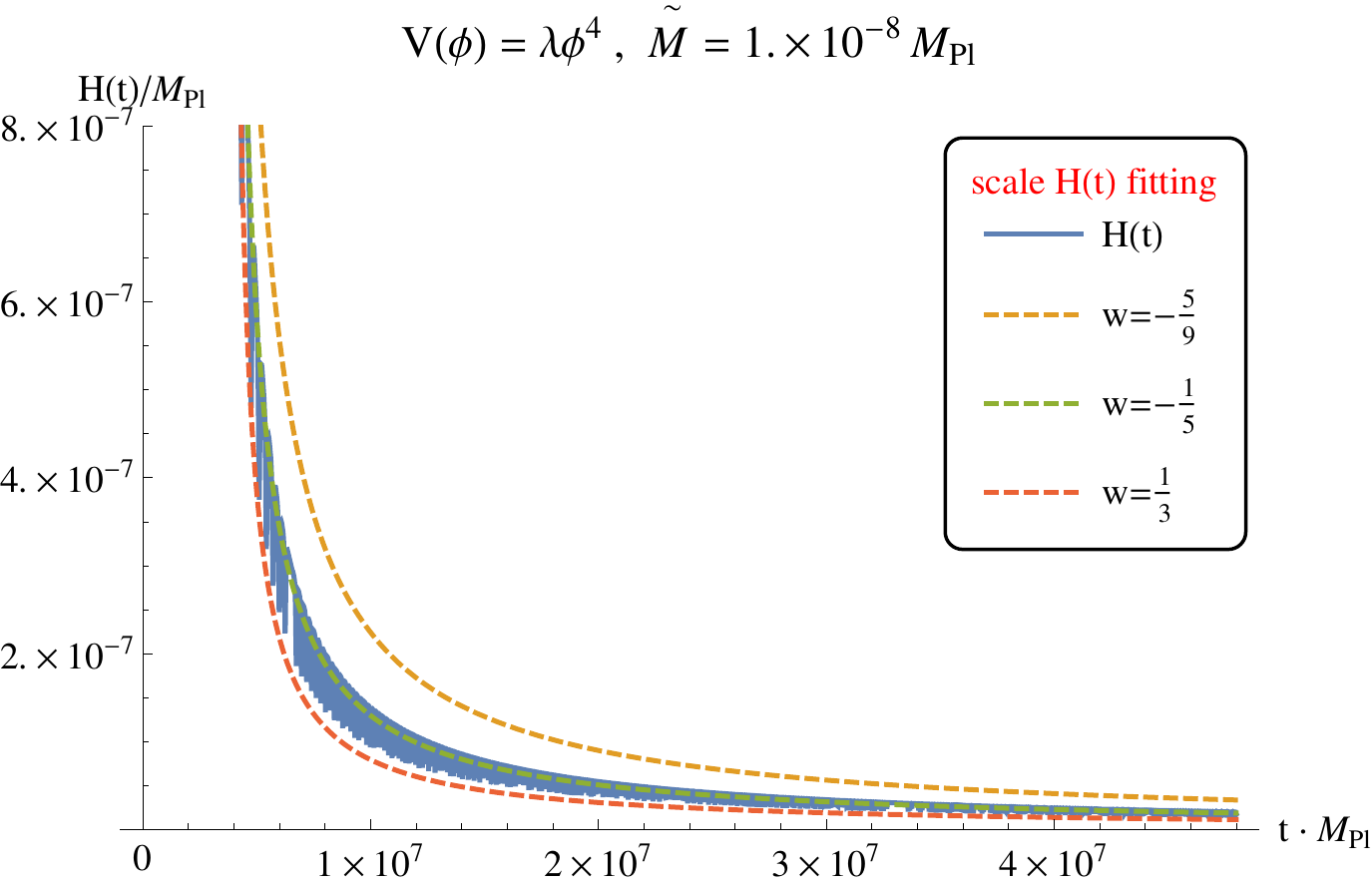} \\
\end{tabular}
\caption{\small{The evolution of the Hubble parameter and fitting curves, for quadratic (left panel) and quartic (right panel) potentials and for $\tilde{M}/M_\text{Pl}=10^{-6}, 10^{-7}, 10^{-8}$. The fitting curves indicate the range of values for the effective averaged EoS, $\bar{w}_\text{reh}$, after the end of the slow-roll inflation $t\gtrsim 7.8 \times 10^{6} M^{-1}_\text{Pl}$.  }}
\end{figure}

The $\bar{H}(t, \bar{w}_\text{reh})$ expression and the non-standard result (\ref{Ht}) of Ref. \cite{Jinno:2013fka, Ema:2015oaa} imply a much different relation for the averaged EoS, $\bar{w}_\text{reh}$, with respect to the shape of the potential
\begin{equation} \label{EoSdc}
\bar{w}_\text{reh(DC)} \sim -\frac{1}{p+1}\,.
\end{equation}
The quadratic potential, $p=2$, yields $\bar{w}_\text{reh}=-1/3$ and the quartic $\bar{w}_\text{reh}=-1/5$ during the oscillating period of the inflaton field. 

Here, we numerically constrain the effective averaged EoS value, $\bar{w}_\text{reh}$, and we find that it deviates significantly from the GR case. In Fig. 6 we plot the actual evolution of the Hubble parameter, $H(t)$, that vividly oscillates and three fitting lines that enclose the $H(t)$. The red-dashed line is the GR case (\ref{potw1}), the green-dashed is the result (\ref{Ht}) of Ref. \cite{Jinno:2013fka, Ema:2015oaa} and the orange-dashed line is the result of Ref. \cite{Sadjadi:2013na}. In particular,  we find for the quadratic and quartic potentials respectively that
\begin{equation}
\frac{2}{3t} \leq  H(t) \lesssim \frac{1}{t}\,,\quad \text{hence} \quad   -1/3 \lesssim \bar{w}_\text{reh} \leq 0\,
\end{equation}
\begin{equation}
\frac{1}{2t} \leq H(t) \lesssim \frac{5}{6t}\,,\quad \text{hence} \quad   -1/5 \lesssim \bar{w}_\text{reh} \leq 1/3\,,
\end{equation}
during the post-inflationary era, here for $t>t_\text{end} \approx 7.8 \times 10^6 M^{-1}_\text{Pl}$. For $\tilde{M} < H(t)$ the line that better fits the averaged value of the Hubble parameter is for $\bar{w}_\text{reh}=-1/(p+1)$ (green-dashed line) and afterwards, when the NMDC becomes negligible, the GR evolution is recovered (red-dashed line). Apparently the smaller the $\tilde{M}$ value is the later the GR evolution is recovered.   This is a striking difference with the minimal GR models which we exploit in this work in order to discriminate the models. In Fig. 8 the purple shaded area corresponds to $-1/3<\bar{w}_\text{reh} \leq 0$ range of values for the NMDC models with the inflaton oscillating about a minimum described by quadratic potential and $-1/5<\bar{w}_\text{reh} \leq 0$ range of values for a minimum described by a quartic potential. These ranges of $\bar{w}_\text{reh}$ values cannot originate from GR models.

It is crucial, however, that the reheating phase does not spoil the inflationary predictions due to the oscillating behavior between positive and negative values of the sound speed squared. We assume that instabilities might be avoided, assumption supported by the results of Ref. \cite{Germani:2015plv} for the new-Higgs inflation.  To this end, the explicit couplings of the inflaton with other degrees of freedom play a crucial role for they control the inflaton lifetime.  The couplings of the inflaton and the gradient instability issue is an interesting study that we leave for a separate work.

The case of the {\itshape exponential} potential is different. Inflation terminates when the condition $\tilde{M}\lesssim H$ is violated. This fact makes the exponential inflation model with NMDC {\itshape safe} from instabilities during the reheating phase. In the post-inflationary era the field does not oscillate; on the contrary it runs away except if a minimum exists in the field space due to extra unspecified dynamics.  During the run-away phase the inflaton field evolves as a minimally coupled field with Einstein gravity. The effective EoS is that of a stiff-fluid, where $p=\rho$, $\bar{w}_\text{reh}=1$ and $\rho \propto a^{-6}$. The energy density of the inflaton field gets redshifted faster than any other energy component and soon becomes subdominant. The radiation produced due to the time varying gravitational field is expected to be a small fraction of the total energy, hence the gravitational waves produced by the inflationary phase ($r=0.16$) will dominate the energy density and the EoS will approach the value $1/3$ \cite{Dodelson:2003vq}. In order for the transition to a radiation dominated universe to take place either the inflaton has to decay very fast, which can happen when the inflaton is coupled to other degrees of freedom \cite{Dalianis:2014nwa}, or another scalar field
 has to dominate the energy density of the universe and finally decay producing the required entropy. Hence, for the exponential potential, we conclude that  the EoS value is larger than zero and less than one, $0<\bar{w}_\text{reh}<1$. A tentative benchmark value is the $\bar{w}_\text{reh}=1/5$ plotted in the Fig. 7 and 8.

The fact the monomial quadratic (quartic) potentials with dominant NMDC yield EoS values roughly $\bar{w}_\text{reh(DC)}\sim -1/3\, (-1/5)$ during the inflaton oscillating period implies that any radiation produced,  e.g via parametric resonances, gets redshifted much faster than the canonical GR case where the quadratic (quartic) potential yields $\bar{w}_\text{reh(GR)}=0 \,(1/3)$. Hence, the approximation that the effective value of the EoS for the NMDC case is mostly determined by the shape of the bottom of the potential where the inflaton oscillates, Eq. (\ref{EoSdc}), is a legitimate assumption.

\subsection{The $N_\text{reh}$ and $T_\text{reh}$ in models with NMDC}

The number of e-folds that take place during the reheating phase are given by the expression (\ref{Nreh1}).
From the previous results we can straightforwardly calculate the logarithm $\ln (\epsilon_*V_*/\rho_\text{end})$ that appears in the Eq. (\ref{Nreh1}). For the monomial potentials the $\epsilon_*$ is given by (\ref{epet}) and can be written in terms of the $\phi_\text{end}$ and $\phi_*$ as
\begin{equation} \label{epga}
\epsilon_* = \gamma^{-1} \left( \frac{\phi_\text{end}}{\phi_*} \right)^{p+2} \,.
\end{equation}
It is also $V_*/\rho_\text{end}= V_*/(\gamma^{-1}V_\text{end})=\gamma\,(\gamma \epsilon_*)^{-p/(p+2)}$. 
Hence, the quantity $(\epsilon_*V_*/\rho_\text{end})$ reads
\begin{equation}
\ln \left( \frac{\epsilon_*V_*} {\rho_\text{end}}\right)\, = \,\ln \left(\gamma \epsilon_*  \right)^\frac{2}{p+2}=\ln \left(\frac{\phi_\text{end}}{\phi_*}\right)^2\, =\, -  \, \frac{2}{(p+2)} \ln \left[1+\frac{2(p+2)}{\gamma \, p} N_* \right]
\end{equation}
and the Eq. (\ref{Nreh1}) for the NMDC scenarios with monomial potentials is recast into
\begin{equation} \label{Ndc}
N_\text{reh} (n_s,p, \bar{w}_\text{reh})\, \simeq \,  \frac{4}{1-3\bar{w}_\text{reh}} \left[ 57.5 - N_*(n_s)  - \frac{1}{2(p+2)} \ln \left(1+\frac{2(p+2)}{\gamma \, p} N_*(n_s) \right) \right]\,,
\end{equation}
where
\begin{equation}
N_*(n_s)=\frac{4(p+1) -\gamma\, p (1-n_s)}{2(p+2)(1-n_s)}\,.
\end{equation}
It is rather interesting that this result can be obtained also from the slow-roll phase correspondence $q = 2p/(p+2)$ from the expression (\ref{Nrehgr}). The small difference comes from the breakdown of the slow-roll approximation, expressed by the $\gamma^{-1}$ factor at the equation (\ref{epga})\,.

As long as the NMDC dominates over the canonical term the $\bar{w}_\text{reh}$ value in Eq. (\ref{Ndc}) is determined by the NMDC dynamics. Afterwards the $\bar{w}_\text{reh}$ approaches its canonical GR value, $w_\text{reh(DC)} \rightarrow w_\text{reh(GR)}$, see Fig. 6. We can split the duration of the reheating stage into the (DC)-stage where the NMDC dominates and the (GR)-stage where the canonical kinetic dominates. We define
\begin{equation}
\left. \left.  N_\text{reh}\, = \, N_\text{reh(DC)}\right|_{\tilde{M} \lesssim H}\, + \, N_\text{reh(GR)} \right|_{\tilde{M} > H}\,
\end{equation}
where
\begin{equation}
 N_\text{reh(DC)} \equiv \frac{1}{3(1+\bar{w}_\text{reh(DC)})} \ln \frac{\rho_\text{end}}{\rho_\text{reh(DC)}}\,, \quad \quad\quad N_\text{reh(GR)} \equiv\frac{1}{3(1+\bar{w}_\text{reh(GR)})} \ln \frac{\rho_\text{reh(DC)}}{\rho_\text{reh(GR)}}\,.
\end{equation}
In the high friction limit, $\tilde{M} \ll H_\text{inf}$, it is reasonable to expect that $N_\text{reh(GR)}$ is restricted since the  inflaton potential parameters $\lambda_p$ are remarkably large. Indeed, the CMB normalized values for the mass of the (non-canonical) inflaton is $m_\phi \gg 10^{-6} M_\text{Pl}$ and the self-coupling is $\lambda_\phi \gg 10^{-14}$. As long as the NMDC effects are dominant the $\lambda_p$ size is "screened" but  the $\lambda_p$ value gradually increases as the Hubble scale decreases.   This implies that the inflaton may decay when $H_\text{} \sim \tilde{M}$, for $\tilde{M} \ll H_\text{inf}$.  If this is the case the averaged value for the EoS can be approximated by the $w_\text{reh(DC)}$.

\begin{figure}
\centering
\begin{tabular}{cc||cc}
{(a)} \includegraphics [scale=.72, angle=0]{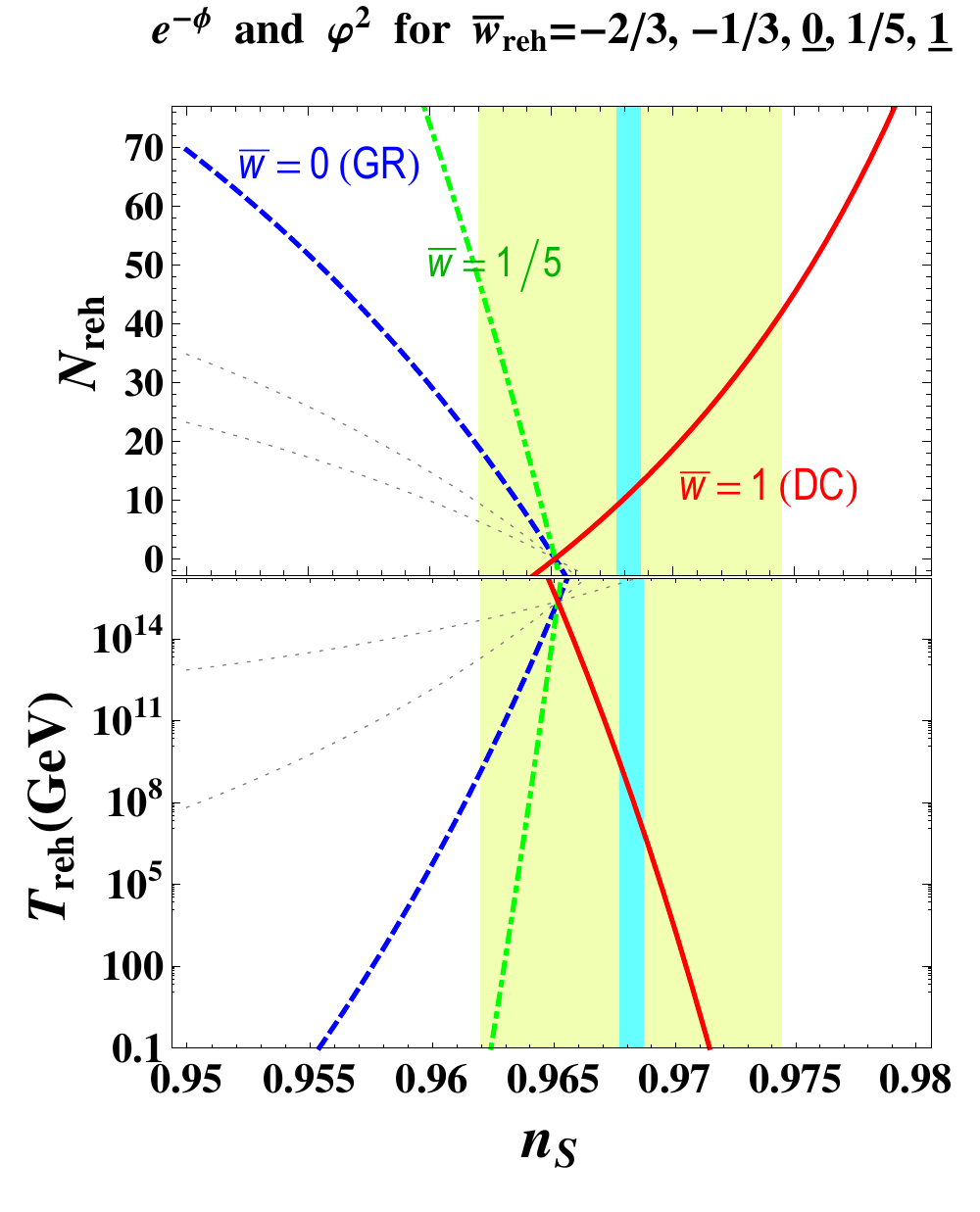} \quad
{(b)} \includegraphics [scale=.72, angle=0]{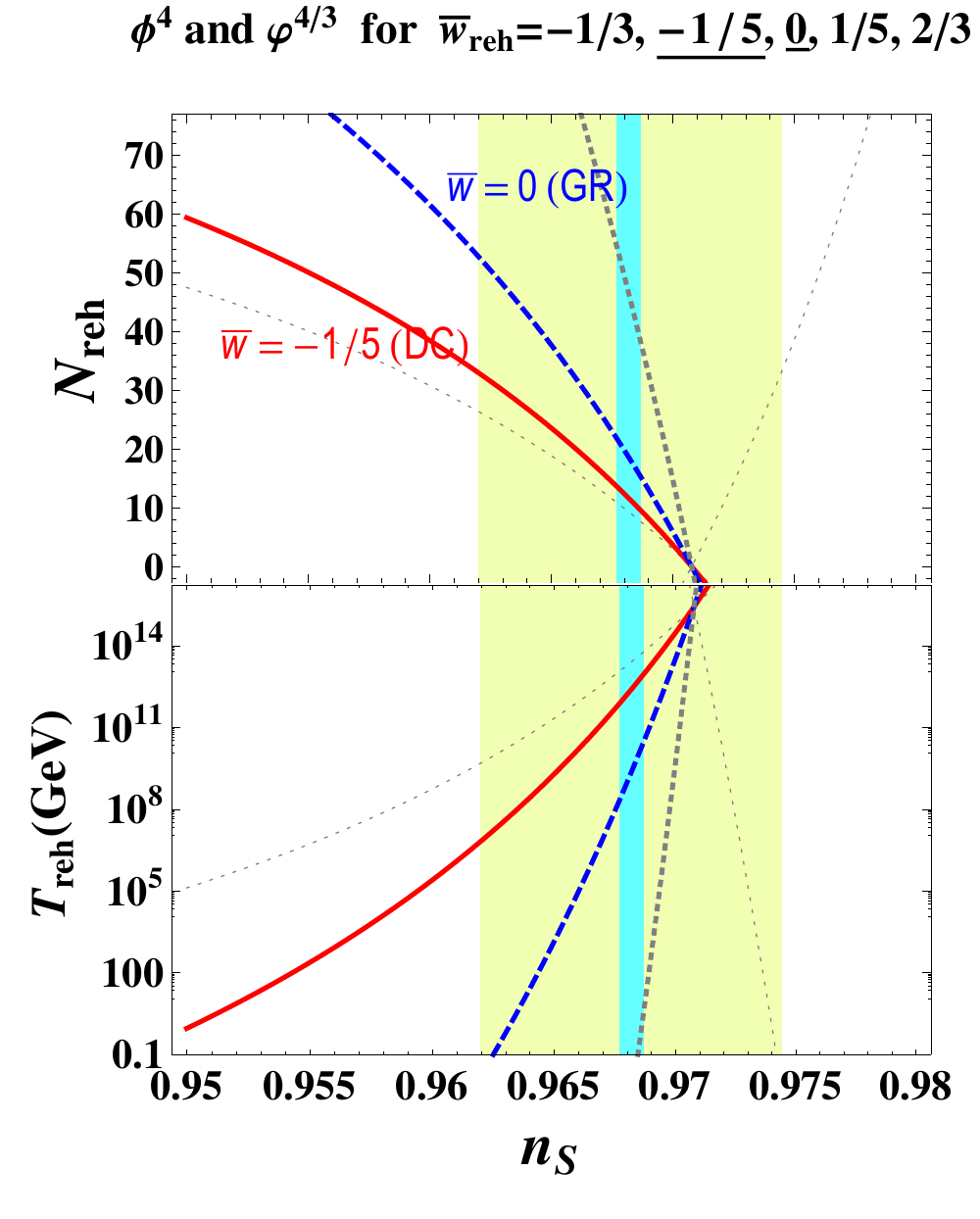}  \\
{(c)} \includegraphics [scale=.72, angle=0]{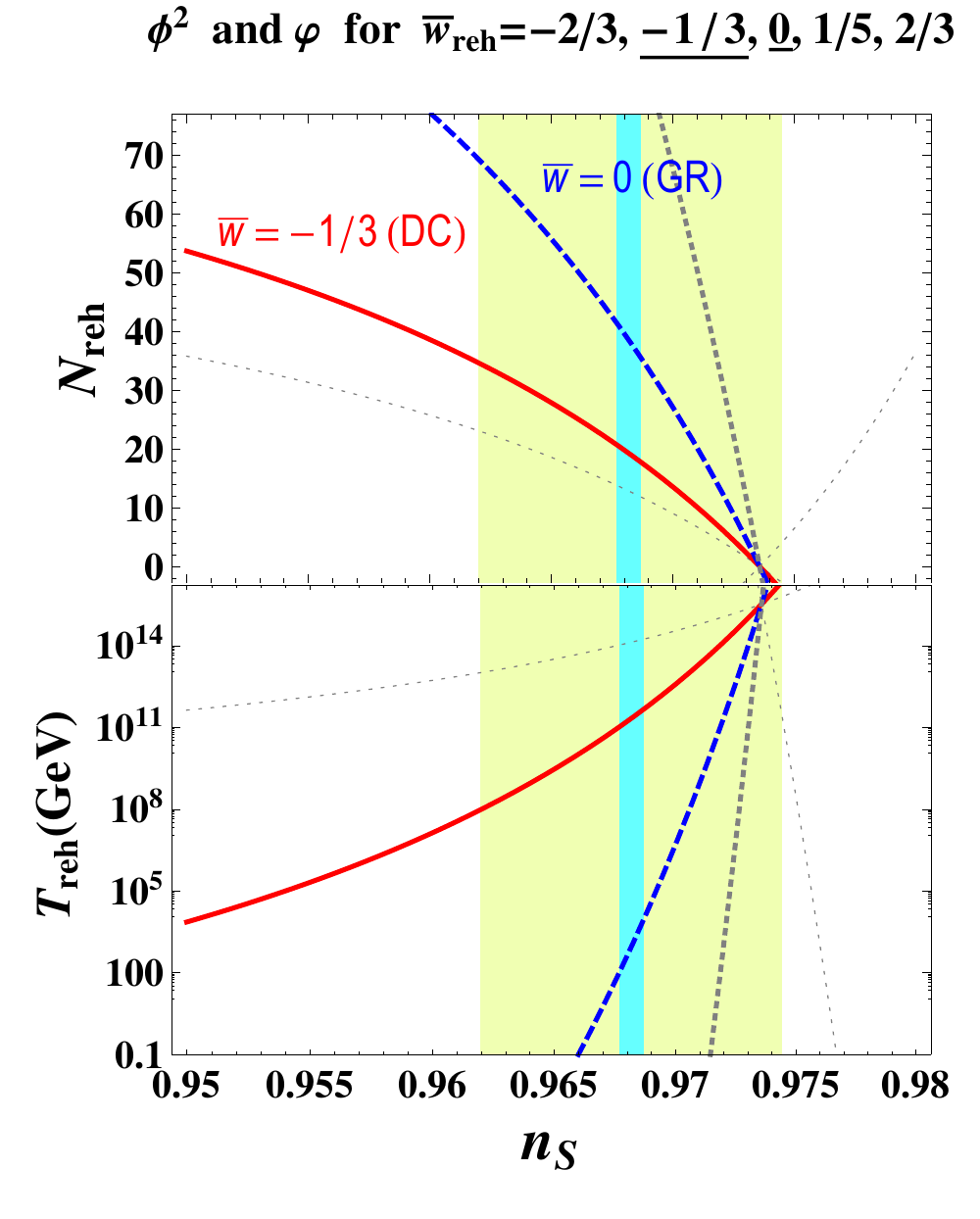} \quad
{(d)} \includegraphics [scale=.72, angle=0]{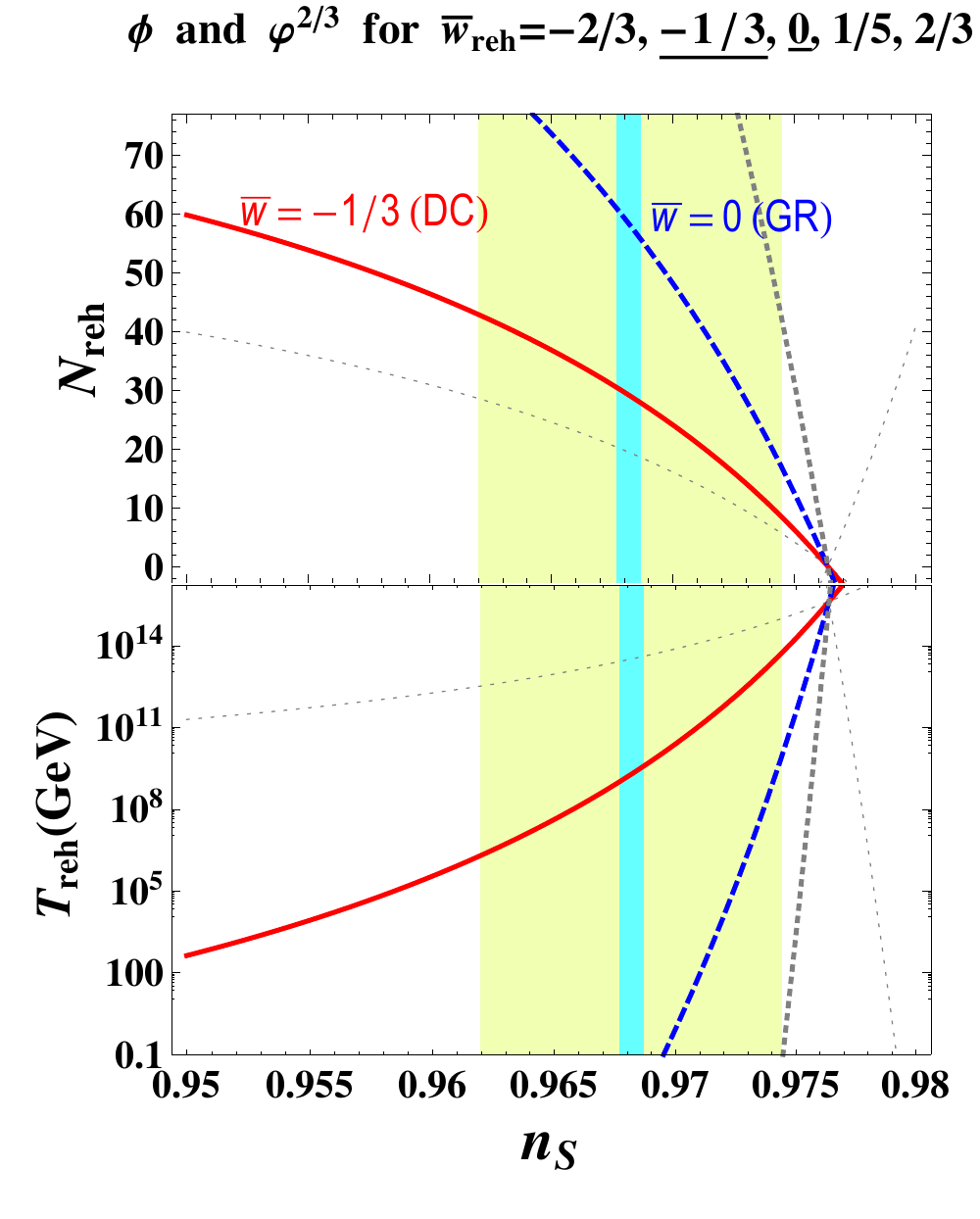}  \\
\end{tabular}
\caption{\small{The plots demonstrate the inflationary predictions for the e-folds number, $N_\text{reh}$, and the reheating temperature $T_\text{reh}$, for the $V(\phi)$-NMDC and $V(\varphi)$-GR models. The wide green band depicts the $1\sigma$ observational uncertainty in the measured $n_s$ value by the Planck satellite, and the narrow cyan band depicts the $n_s$ central value. Ignoring the small correction due to the $\gamma$ factor,
the predictions coincide modulo the unknown $\bar{w}_\text{reh}$ parameter value during the reheating phase.  The curves in red are the benchmark values for the NMDC models and in blue-dashed for the GR models.
In thick dotted-gray is for $\bar{w}_\text{reh}=1/5$, value indicated by thermalization scenarios, and in thin dotted-gray other values for the $\bar{w}_\text{reh}$ parameter are shown for comparison.
}}
\end{figure}

The CMB normalization constrains the NMDC mass scale to be  $\tilde{M}\gtrsim 10^{-8-14} M_\text{Pl}$ and the evolution of the Hubble scale indicates that $-1/3 \lesssim \bar{w}_\text{reh} \leq 0$, see Fig. 6. This range for $\tilde{M}$ and $\bar{w}_\text{reh}$ values implies that the reheating period with dominant NMDC effects is bounded between the values 
\begin{equation} \label{nest}
 1 \, \lesssim N_\text{reh(DC)} \, \lesssim  10 \,.
\end{equation}
The exact value depends on the  power $p$, the $\tilde{M}$ and the full couplings of the inflaton to other degrees of freedom.

In addition, for the monomial potentials the reheating temperature (\ref{Treh1}) can be written in terms of the parameters $p$, $\bar{w}$ and the observable quantity $n_s$,
\begin{equation} \label{Tdc}
T_\text{reh}(n_s, p, \bar{w}_\text{reh})= \left(\frac{1}{\gamma}\right)^{1/4} \lambda^{1/4}_p \left(\frac{p}{\sqrt{3\lambda_p}}\, \tilde{M}^2M^p_\text{Pl}  \right)^\frac{p}{4p+8}  \left(\frac{30}{\pi^2 g_*}\right)^{1/4} M^\frac{4-p}{4}_\text{Pl} e^{-\frac34(1+\bar{w}_\text{reh}) N_\text{reh}(n_s)}.
\end{equation}
Assuming no violation of the null-energy condition for the averaged EoS value, the maximum reheating temperature, $T_\text{max}=T_\text{max}(p, \lambda_p, \tilde{M})$, is the coefficient in front of the exponential in Eq. (\ref{Tdc}). For $N_\text{reh}>0$ values the reheating temperature decreases exponentially.

If the NMDC is effective till the time of the perturbative inflaton decay, that is $N_\text{reh(GR)} \rightarrow 0$, 
a case expected for $\tilde{M}\ll H_\text{inf}$, then the estimation (\ref{nest}) implies that the reheating temperature, given by Eq. (\ref{Tdc}), lies in the range
\begin{equation} \label{Treh}
 10^{-3} \, T_\text{max} \, \lesssim \,T_\text{reh} \, \lesssim \, T_\text{max}.
\end{equation}
If the $N_\text{reh(GR)}$ is not negligible then the $\bar{w}_\text{reh}$ departs from the benchmark NMDC value (\ref{EoSdc}) but, still, it is expected to be less than the benchmark GR value (\ref{potw1}). The purple shaded region in Fig. 8 elucidates this "hybrid" case where the NMDC and the canonical term are comparable during the reheating stage.
The above discussion suggests that the unknown mass scale $\tilde{M}$ can be constrained by the (indirect) measurement of the reheating temperature $T_\text{reh}$ and the reheating duration $N_\text{reh}$.

The $N_\text{reh}$ is plotted in Fig. 7 and the $T_\text{reh}$ in the Fig. 7 and 8.

\begin{figure}
\centering
\begin{tabular}{cc||cc}
\includegraphics [scale=.42, angle=0]{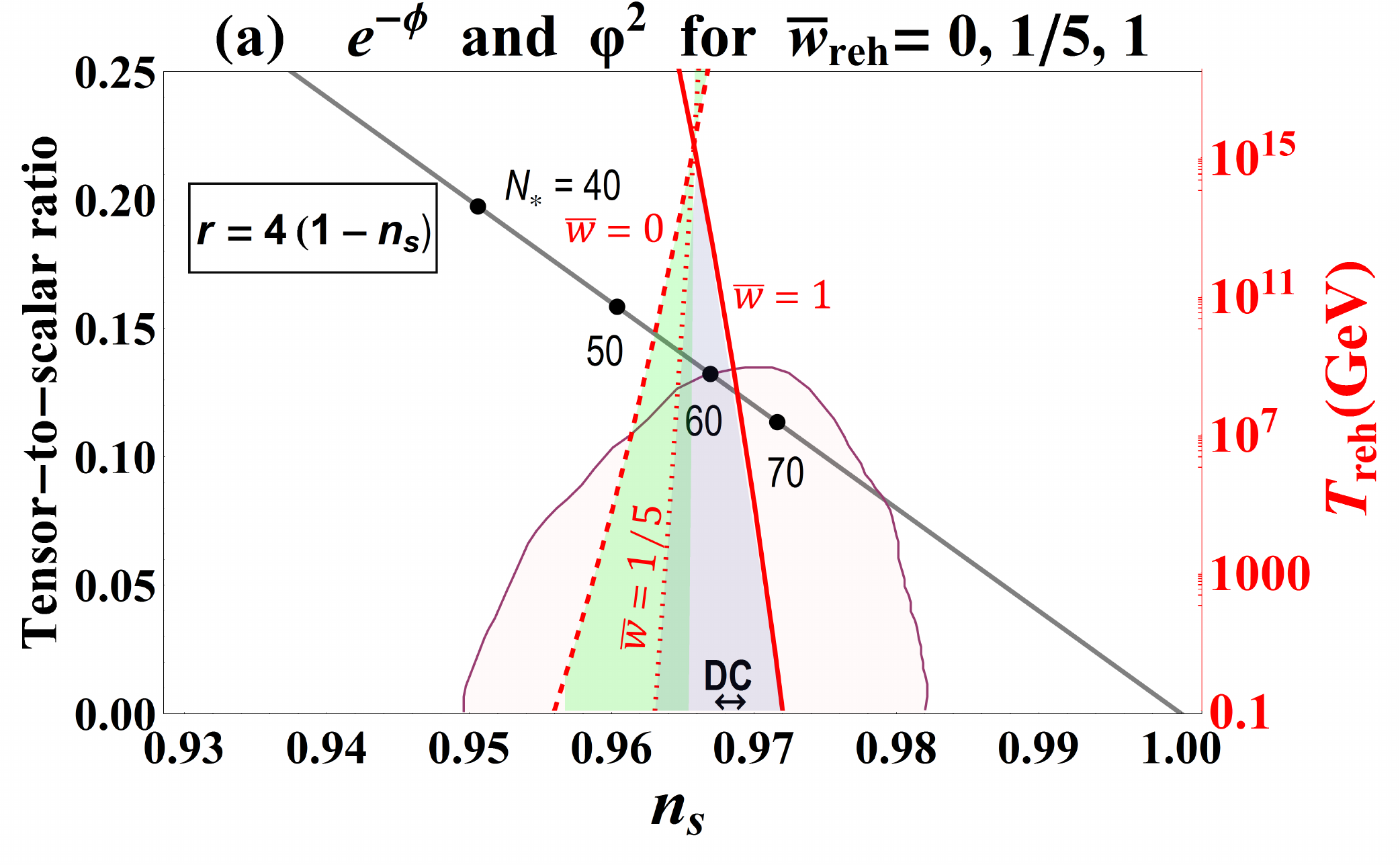} \quad
\includegraphics [scale=.42, angle=0]{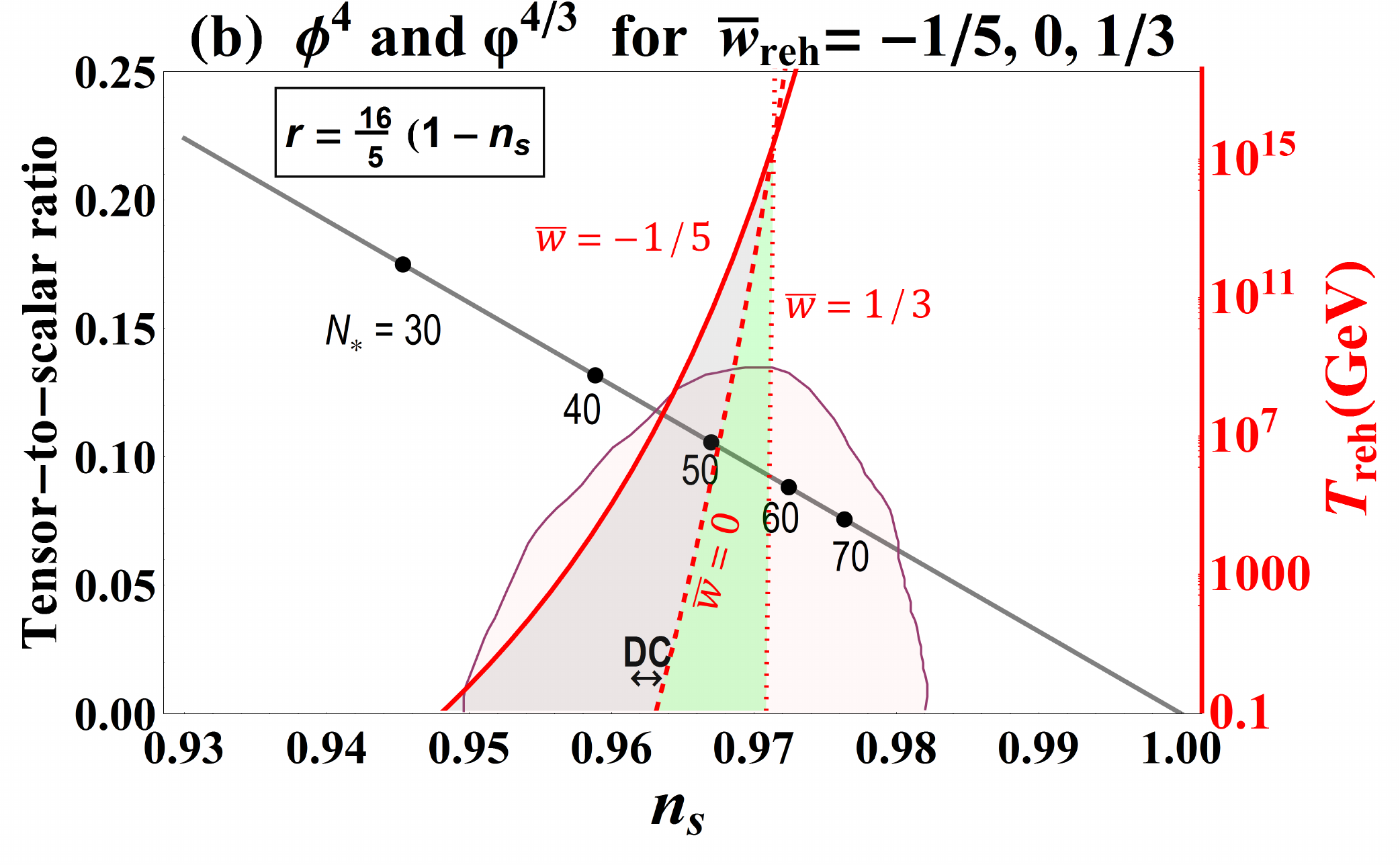}  \\
 \includegraphics [scale=.42, angle=0]{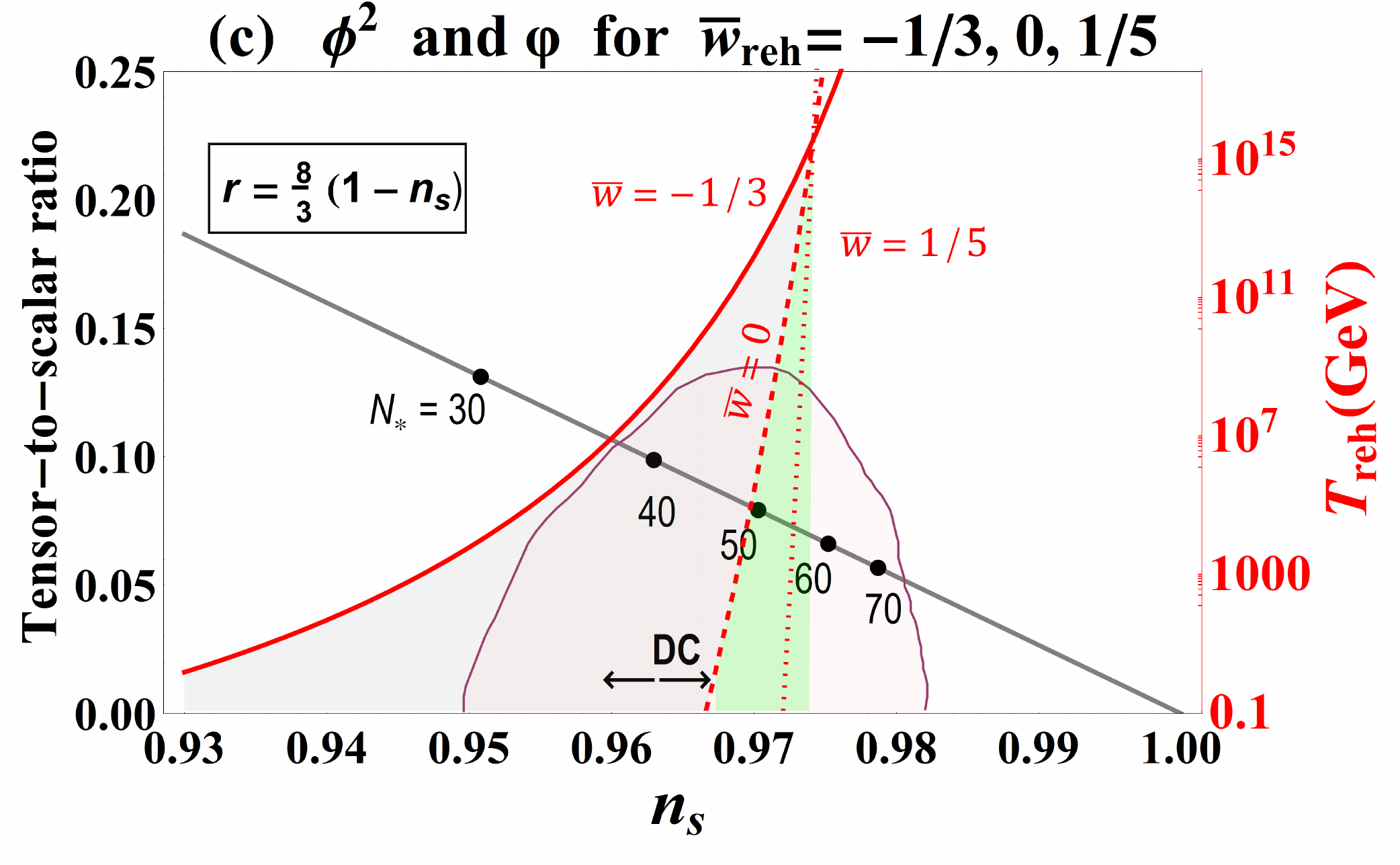} \quad
 \includegraphics [scale=.42, angle=0]{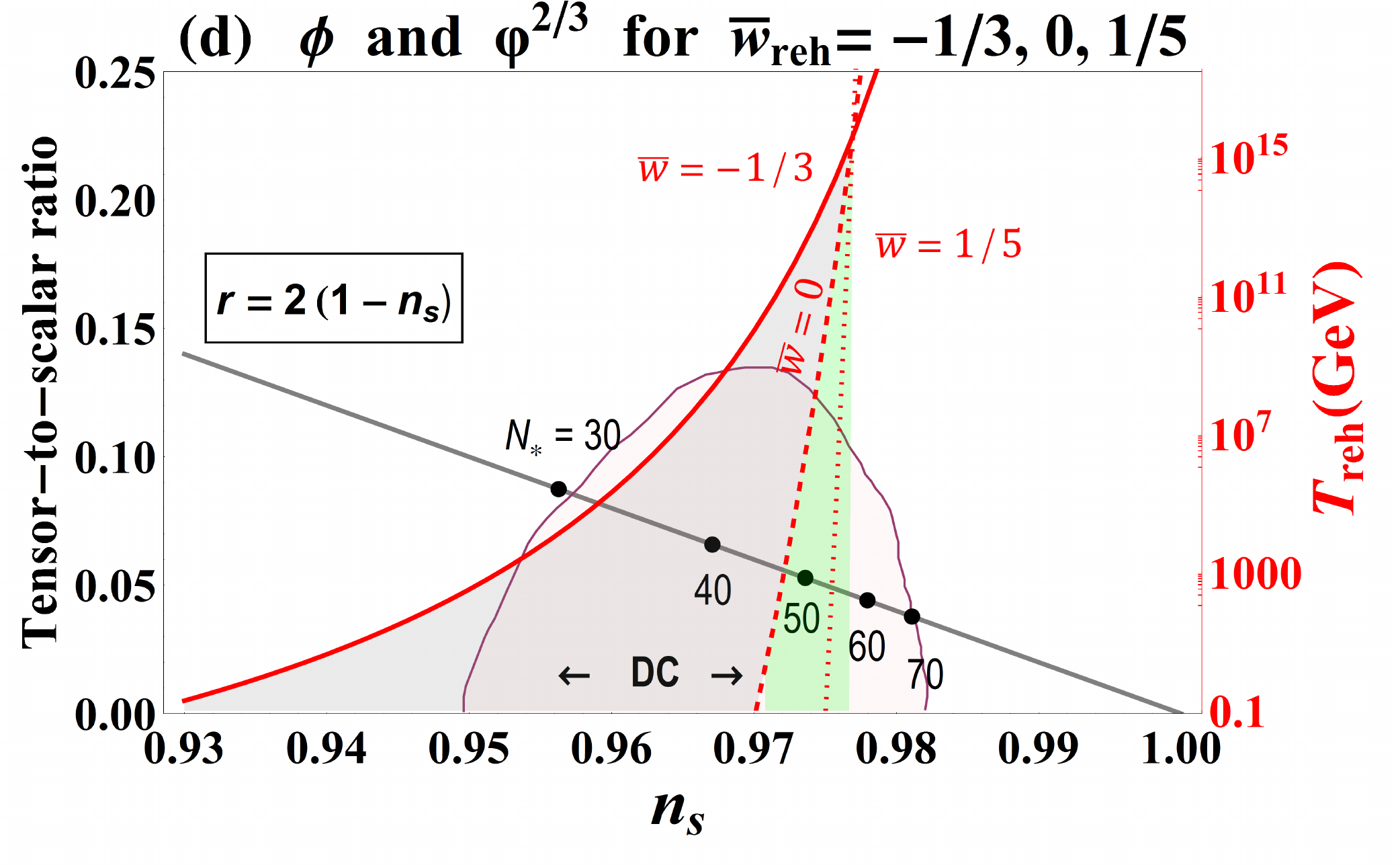}  \\
\end{tabular}
\caption{\small{ The plots have two different vertical axes: the tensor-to-scalar-ratio $r$ and the $T_\text{reh}$. In the $(n_s, r)$ axes-system we plot the marginalized joint $95\%$ CL observational contour for the  $n_s$ and the $r_{0.002}$ from  {\itshape Planck} TT$\_\text{lowP}$ $2015$; and the theoretical predictions for the $r=r(n_s)$ in straight blue line. In the $(n_s, T_\text{reh})$ axes-system we plot the predicted reheating temperature $T_\text{reh}$  in red-dashed curves for the GR models and red-solid curves for the NMDC models where benchmark EoS values are used. The purple shaded area gives the $T_\text{reh}$ for non-benchmark EoS values that are broadly admitted by the NMDC models, and the green shaded area the $T_\text{reh}$ for EoS values $0\leq\bar{w}_\text{reh} \leq 1/3$ admitted by the GR models.
The inflationary minimum is assumed to be approximately quadratic, except for the panel (a) where the NMDC-$\phi$ runs away right after inflation and the (b) where a quartic potential about the minimum is considered. In the $\Delta n_s$ values where the NMDC and the GR predictions overlap the NMDC models predict larger reheating temperatures.  On the other hand, the $\Delta n_s$ values indicated by the arrows labeled "DC" cannot be naturally explained by the GR models and favor the presence of the NMDC.
}}
\end{figure}

\section{Model selection}

In Fig. 1 the spectral tilt $n_s$ and the tensor-to-scalar-ratio  $r$ predictions of some basic representative inflationary models with canonical kinetic term in GR and with NMDC are shown.  When the NMDC accounts for a negligible correction to the kinetic term, i.e. $\tilde{M} \gg H_\text{inf}$ then the predictions are indistinguishable with the standard GR predictions (GR-limit). As the value of the $\tilde{M}$ decreases the NMDC effects start becoming manifest with predictions that depart from those of GR. In the intermediate region, $\tilde{M} \sim H_\text{inf}$ the $r=r(n_s)$ lines lie in the lightyellow area of ($n_s, r$) plane, Fig. 2 and 3. The interesting feature of the intermediate region of the NMDC inflationary models is that the predicted $(n_s, r)$ values may be quite unusual.

The most interesting case is the high friction limit where the NMDC dominates, $\tilde{M} \ll H_\text{inf}$, because steep potentials can inflate the universe and, in addition, be compatible with the latest observational constraints.
During the slow-roll phase and for $\tilde{M} \ll H_\text{end}$ the NMDC system enjoys a dual description in terms of a minimally coupled field in GR. As a result, the NMDC predictions $(n_s, r)$ fall together with the $(n_s, r)$ predictions of standard GR inflationary models, such as the axion monodromy models.
This degeneracy in the inflationary predictions obscures the model selection process and prevents the observations to conclude
against or in favor of this type of modified gravity theories. However, the theoretical degeneracy breaks when the reheating stage is taken into account because the NMDC models predict $\bar{w}_\text{reh}$ values different than those of their de-Sitter GR duals.

In Fig. 7 we plotted the length of the reheating stage, $N_\text{reh}$, and the reheating temperature, $T_\text{reh}$.
The green shaded band corresponds to the $1\sigma$ bounds on spectral index of curvature perturbations $n_s$ from the {\itshape Planck} full mission temperature and polarization data (2015) on large angular scales measure \cite{Ade:2015lrj}, $n_s=0.968 \pm 0.006$ (68\% CL). The narrow cyan band about the central value has a width $\Delta n_s=10^{-3}$, accuracy that is expected to be achieved by future CMB experiments. The plots are truncated at the value $T_\text{reh} <0.1$ GeV, in order to make BBN, dark matter production and baryogenesis possible. In the plots the lines intersect at a point which corresponds to instant reheating; beyond that point the $N_\text{reh}$ is negative.
The predictions of the GR models, depicted in dashed and thick-dotted lines, are given by the expressions (\ref{Nrehgr}) for the $N_\text{reh}$ and (\ref{Tgr}) for the $T_\text{reh}$. The predictions of the NMDC models, depicted in solid lines 
 are given by the expressions (\ref{Ndc}) for the $N_\text{reh}$ and (\ref{Tdc}) for the $T_\text{reh}$. The GR and NMDC predictions $N_\text{reh}$, $T_\text{reh}$ coincide when the correspondence (\ref{corre}) between the dual models is applied and in the limit $\gamma\rightarrow 1$. The main difference between the GR and the NMDC models is the value of the effective EoS parameter after inflation and practically this is the distinctive feature for the predicted $N_\text{reh}$ and $T_\text{reh}$ values.

In Fig. 8 we plotted the $T_\text{reh}(n_s)$ predictions against the observationally constrained $(n_s,r)$ contour from {\itshape Planck} 2015.  The plots are two-scaled, for each $n_s$ value the predicted $r$ and $T_\text{reh}$ values are shown. The $r=r(n_s)$ predictions are given by the expressions (\ref{A}) for the GR models and (\ref{ns2}) for the NMDC. 
In the plots, the small correction due to $\gamma$ factor is not visible, therefore the degeneracy between the "dual" GR and NMDC models is practically broken by the different value of the effective EoS parameter after inflation.
\\
\\
Below, we compare the $N_\text{reh}$ and $T_\text{reh}$ predictions of the GR and NMDC models that yield the same $r=r(n_s)$ relation using the the marginalized joint $95\%$ CL observational contour for the  $n_s$ and the $r_{0.002}$ from  {\itshape Planck} TT$\_\text{lowP}$ $2015$. 
Benchmark values for the $\bar{w}_\text{reh}$ are taken. Here, we introduce the subscript GR in the potential for the canonical field for clarity.
\\
\begin{description}
\item [(a)] \underline{$r=4(1-n_s)$} : $\,\, V_\text{}(\phi)\propto e^{-\phi}$ and $V_\text{GR}(\varphi)\propto \varphi^2$
\end{description}
The {\itshape Planck} analysis selects the values $0.967<n_s<0.979$ and $0.082<r<0.131$.  
The $\varphi^2$ GR model is mostly ruled out by the data since the accepted $N_*>60$ values are realized only for unnatural $\bar{w}_\text{reh}$ values. The NMDC can naturally save the $r=4(1-n_s)$ line in the $(r, n_s)$ contour thanks to the larger $\bar{w}_\text{reh}$ values predicted, albeit extra dynamics are required for a prompt inflaton decay. The exponential potential with NMDC can be compatible with the data only if the reheating temperature is very large, $T_\text{reh}>10^{13}$ GeV.
\\
\begin{description}
\item [(b)] \underline{$r=(16/5)(1-n_s)$} : $\,\, V_\text{}(\phi)\propto \phi^4$ and $V_\text{GR}(\varphi)\propto \varphi^{4/3}$
\end{description}
The {\itshape Planck} analysis selects the values $0.963<n_s<0.98$ and $0.063<r<0.117$.
Both models can fit well the data. For a given $n_s$ value the NMDC always predicts higher reheating temperatures and a shorter reheating stage.   For $N_\text{reh}\sim 10$ and benchmark $\bar{w}_\text{reh(DC)}$ values we take $n_s \sim 0.968$ and $T_\text{reh}\sim 10^{12}$ GeV, about three orders of magnitude above the expected GR reheating temperature value.
If values $n_s<0.964$ are selected by the future experiments the  $r=(16/5)(1-n_s)$ line remains viable only thanks to the NMDC. If $n_s>0.971$ neither the GR nor the NMDC models can naturally explain the $r=(8/3)(1-n_s)$ line. \\
\begin{description}
\item [(c)] \underline{$r=(8/3)(1-n_s)$} : $\,\,V_\text{}(\phi)\propto \phi^2$ and $V_\text{GR}(\varphi)\propto \varphi$
\end{description}
The {\itshape Planck} analysis selects the values $0.96<n_s<0.98$ and $0.051< r <0.105$.
Both models can fit the data with the NMDC model predicting a shorter reheating stage and reheating temperature order of magnitudes larger than the GR .
The GR model implies a prolonged reheating period for the lower values of $n_s$.
For $N_\text{reh}\sim 10$ and benchmark $\bar{w}_\text{reh(DC)}$ values we take $n_s \sim 0.97$ and $T_\text{reh}\sim 10^{13}$ GeV, about four orders of magnitude above the expected GR value.
If the observed $n_s$ value is constrained to values less than $n_s <0.967$ the $r=(8/3)(1-n_s)$ line selects the NMDC, whereas the GR model is ruled out.
On the contrary, if $n_s>0.972$ neither the GR nor the NMDC models can naturally explain the $r=(8/3)(1-n_s)$ line.\\
\begin{description}
\item [(d)] \underline{$r=2 (1-n_s)$} : $\,\,V_\text{}(\phi)\propto \phi$ and $V_\text{GR}(\varphi) \propto \varphi^{2/3}$
\end{description}
The {\itshape Planck} analysis selects the values $0.957<n_s<0.981$ and $0.037<r<0.085$.
Also here, the NMDC model predicts a shorter reheating stage and larger values for the reheating temperature by many orders of magnitudes for the greatest part of the parameter space. The GR model implies a rather prolonged reheating period for the lower values of $n_s$. For $N_\text{reh}\sim 10$ and benchmark $\bar{w}_\text{reh(DC)}$ values we take $n_s \sim 0.974$ and $T_\text{reh}\sim 10^{13}$ GeV, about four orders of magnitude above the expected GR value.
If the observed $n_s$ is constrained to values less than $n_s <0.97$ the $r=2 (1-n_s)$ line  selects the NMDC and disfavors the GR-model. If $n_s>0.978$ neither the GR nor the NMDC models can naturally save the $r=(8/3)(1-n_s)$ line. \\

Apart from the differences in the reheating temperature the NMDC models have additional differences with the GR models, that may play a non-negligible role during the reheating period. The CMB normalization yields a mass for the inflaton field much larger than the standard inflationary models. If the inflaton decay rate is suppressed then it may decay when $H_\text{} \lesssim \tilde{M}\ll H_\text{inf}  $ which implies that the perturbative production of very heavy particles, $m_\phi \gg 10^{13}$ GeV, is possible. Such heavy particles are welcome in scenarios associated with the leptogenesis mechanism, e.g. the right-handed neutrinos. 

On the observational side, the recent and the forthcoming experimental advances will place stringent bound on the predictions of inflationary models.
Future measurements of the $n_s$, $r$ and the reheating temperature  by experiments such as EUCLID \cite{Amendola:2012ys}, PRISM \cite{Andre:2013afa}, LiteBIRD \cite{Matsumura:2013aja} and the DECIGO \cite{Kuroyanagi:2014qza} can provide a decisive test for the shape of the inflaton potential and the Horndeski-type theories. On the theoretical side, a full exploration of the postinflationary validity of the perturbation theory for the NMDC models along the lines of Ref. \cite{Ema:2015oaa, Germani:2015plv, Myung:2016twf} should be carried out.

\section{Conclusions}

In this paper we discuss the reheating predictions in single field primordial inflation described by a Lagrangian with non-minimal derivative coupling (NMDC) to the Einstein tensor. Models with NMDC account for a new branch of inflation model building with attractive features as the extension of the parameter space that implements inflation and the suppression of the tensor-to-scalar ratio.
During the slow-roll phase 
 the dynamics of models with dominant the NMDC term become practically indistinguishable from GR models, that can be seen as de-Sitter duals.
We show that the degeneracy in the inflationary predictions, $r=r(n_s)$, with GR models can break when the postinflationary reheating period is taken into account.

The way the universe evolves during the reheating phase affects the predictions for inflation because it determines how the observed CMB scales are mapped back to the inflationary epoch.
Our ignorance for the reheating stage can be parametrized in terms of the averaged equation of state (EoS), the duration  of the reheating stage $N_\text{reh}$ and the final temperature $T_\text{reh}$.  We derive the new expressions for the $N_\text{reh}$ and  $T_\text{reh}$ in terms of the observe quantity $n_s$ for the NMDC models.
Under the reasonable assumption that the averaged EoS after inflation is determined mostly by the oscillating behavior of the inflaton field itself about the minimum of the potential, we perform a comparative study of the NMDC and GR models. For the GR models we have taken the benchmark EoS values $0$, assuming approximately a quadratic potential for low field values, and $1/5$ for scenarios where the thermalization process takes place efficiently. For the NMDC models the central benchmark EoS values are found to be negative, with the EoS range of values being $-1/3\lesssim \bar{w}_\text{reh} \leq 0$ for the $\phi^2$ or $-1/5 \lesssim \bar{w}_\text{reh} \leq 1/3$ for the $\phi^4$ inflation, due to the peculiar reheating dynamics of this class of models.
Thus, the reheating period can be a powerful discriminant of the NMDC scenario. Some crucial reheating issues, such as the possible breakdown of the linear regime during the postinflationary evolution, remain open and deserve further investigation.

Our main results are summarized in Fig. 7  and 8.
The general result is that the NMDC models predict much higher values for the reheating temperature and the NMDC models are compatible with a larger part of the observationally constrained $(n_s, r)$ plane than their GR duals.
The current and forthcoming satellite observations can pin-down a range for the $n_s$ value which translates into a range for
the e-folds number $N_*$ and equivalently into reheating temperature range. For a given model the CMB observations can indicate the effective EoS during reheating (or the reheating temperature directly \cite{Kuroyanagi:2014qza}) and therefore possibly test the presence of the NMDC in the inflaton field dynamics.  The size of the NMDC scale $\tilde{M}$ is also possible,  in principle, to  be probed by the measurement of the reheating temperature; complementary theoretical studies regarding the short-scale instabilities should constrain further the allowed $\tilde{M}$ values.

Our analysis is model dependent at certain points, e.g. the effective averaged EoS and the IR completion of the models.  We have chosen benchmark values, indicated by the shape of the potential and the reheating dynamics, in order the comparison between the models to be carried out; nevertheless, intermediate values for the $\bar{w}_\text{reh(DC)}$ and $\bar{w}_\text{reh(GR)}$ were considered in the plots as well.  Although our results narrow the parameter space where the models are practically indistinguishable,
 they are not conclusive due to the theoretical uncertainties on the reheating stage and the current observational limitations.
Upcoming CMB experiments promise to reduce the $\delta n_s$ and $\delta r$ uncertainty to ${\cal O}(10^{-2}-10^{-3})$ level, making the observational discrimination between different inflationary mechanisms possible.

\section*{Acknowledgments}
We would like to thank Shinji Tsujikawa and Yuki Watanabe  for correspondence.

\vspace*{.5cm}
\noindent

\appendix

\section{Correspondence between NMDC and GR dynamics during inflation}

\subsection{The potentials for an inflaton with NMDC and a canonical inflaton with GR}

\begin{figure}
\centering
\begin{tabular}{cc}
{(a)} \includegraphics [scale=.35, angle=0]{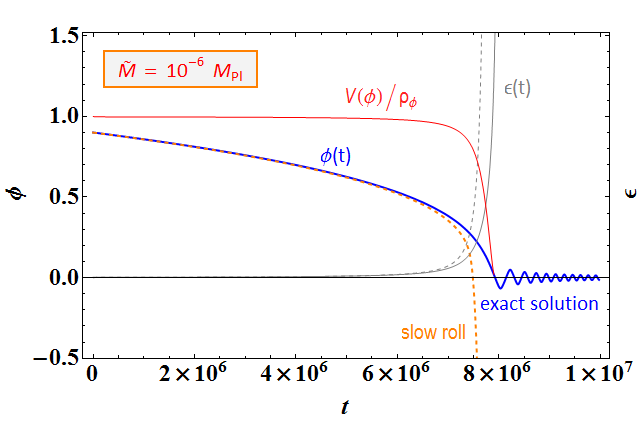} \quad
{(b)} \includegraphics [scale=.35, angle=0]{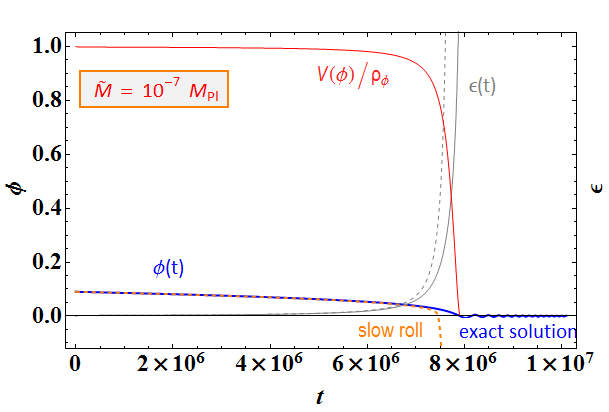}  \\
\end{tabular}
\caption{\small{The exact evolution of the field $\phi(t)$ (continuous blue line) versus the slow-roll approximation (orange-dashed line) for an inflaton with NMDC and quadratic potential. The end of inflation and the breakdown of the slow-roll approximation is shown by the evolution of the slow-roll parameter $\epsilon=-\dot{H}/H^2$ (continuous gray line); the slow-roll approximated value $\epsilon = \epsilon_V/ (3H^2 \tilde{M}^{-2})$ is also shown (gray-dashed line). The scaling with time of the ratio of the potential energy over the energy density, $ V(\phi)/\rho_\phi$, is additionally displayed (red line) to indicate the $\gamma$ factor value at the end of inflation. The left panel corresponds to $\tilde{M}=10^{-6} M_\text{Pl}$ and the right to $\tilde{M}=10^{-7} M_\text{Pl}$. The figures demonstrate that the slow-roll approximation (\ref{syst}) describes very well the actual evolution during the slow-roll period. Planck mass and Planck time units are used.
}}
\end{figure}
During the slow-roll regime and in the high friction limit, $H^2\gg \tilde{M}^2$, the EOM of an inflaton field $\phi$ with non-minimal derivative coupling to the Einstein reads approximately
\begin{equation}
3H\dot{\phi} \left(1+3\frac{H^2}{\tilde{M^2}}\right) + V'_\text{}(\phi) \approx 0\,
\end{equation}
Together with the Friedmann equation, the dynamics of the slowly rolling field are well described, see Fig. 9, by the system of the
truncated equations:
\begin{align} \label{syst}
H^2\simeq \frac{V_\text{}(\phi)}{3M^2_\text{Pl}}\,,\quad\quad 3H\dot{\phi} \simeq - \frac{\epsilon}{\epsilon_V}V'(\phi)\,.
\end{align}
There is {\itshape generic} transformation of the form
\begin{align}
\varphi=g(\phi), \quad\quad V_\text{GR}(\varphi)=V[g^{-1}(\varphi)]
\end{align}
such that the above system of equations (\ref{syst}) is recast into
\begin{equation} \label{msyst}
H^2\simeq \frac{V_\text{GR}(\varphi)}{3M^2_\text{Pl}},\quad\quad 3H\dot{\varphi} \simeq -V'_\text{GR}(\varphi)\,,
\end{equation}
where $V_\text{GR}(\varphi)$ a potential for the field $\varphi$ minimally coupled to gravity (here, for clarity, we explicitly use the the subscript GR). After straightforward calculations the EOM of (\ref{msyst}) is written in terms of the $\phi$ field as
\begin{align}
3H\dot{\phi} \simeq -\frac{V'(\phi)}{[g'(\phi)]^2}\,,
\end{align}
where the prime denotes derivative with respect to the argument field.  This equation is equivalent to the EOM of the system (\ref{syst}) if $[g'(\phi)]^2=\epsilon_V/\epsilon$ or 
\begin{align}
g'(\phi)= \frac{V^{1/2}}{M_\text{Pl}\tilde{M}}\,.
\end{align}
Therefore the new field $\varphi$ reads in terms of the field $\phi$
\begin{equation} \label{trans}
\varphi=\int\frac{V^{1/2}}{M_\text{Pl}\tilde{M}}\,d\phi\,.
\end{equation}
For the exponential potential  $V=V_0e^{-\lambda_e \phi/M_\text{Pl}}$ we take $\varphi= -{2V^{1/2}_0}/({\lambda_e \tilde{M}}) \times e^{-\lambda_e \phi/(2M_\text{Pl}})$.
and the inverse function, $g^{-1}(\varphi)=\phi=-(2M_P/\lambda_e)$ $\times$ $\ln(-\lambda_e \tilde{M}\varphi/2V^{1/2}_0)$. It follows that  the potential $V_\text{GR}$ for the minimally coupled $\varphi$ field reads
\begin{align}
V_\text{GR}(\varphi)=V[g^{-1}(\varphi)]=\frac12\frac{\lambda^2_e\tilde{M}^2}{2}\varphi^2 \equiv \frac12 m^2_\varphi \varphi^2\,.
\end{align}
Let us consider the monomial potentials
\begin{equation}
V_\text{}(\phi)= \lambda_p M^{4-p}_\text{Pl} \phi^p\,.
\end{equation}
According to the transformation (\ref{trans}), $\varphi=\int g'(\phi)d\phi$, the field $\varphi$  reads
\begin{equation}
\varphi = \frac{2}{p+2} \frac{\phi^{p/2+1}}{\lambda^{-1/2}_p M_\text{Pl}^{p/2-1}\tilde{M}}\,,
\end{equation}
and it appears to be minimally coupled to gravity during the inflationary phase. Its evolution is governed by the potential $V_\text{GR}(\varphi)=V[g^{-1}(\varphi)]$
where
\begin{equation} \label{vmm}
V_\text{GR}(\varphi) = \lambda_pM^{4-p}_\text{Pl}\left(\frac{p+2}{2} \lambda^{-1/2}_p M_\text{Pl}^{p/2-1}\tilde{M}\,\varphi \right)^{2p/(p+2)}\,.
\end{equation}
Apparently, there is direct correspondence between the potential $V_\text{}(\phi)$ for the NMDC inflaton and the $V_\text{GR}(\varphi)$ for the minimally coupled inflaton and GR:
\begin{equation} \label{arr}
V\propto \phi^p \quad \longleftrightarrow \quad V_\text{GR}\propto \varphi^{\frac{2p}{p+2}}\,.
\end{equation}
Let us look into specific examples, starting from the quartic Higgs-like potential. We find that
\begin{equation}
V_\text{}(\phi)=\lambda_4 \phi^4\quad \longleftrightarrow \quad V_\text{GR} (\varphi)=\lambda_4^{1/3}(3M_\text{Pl}\tilde{M})^{4/3} \varphi^{4/3}\,,
\end{equation}
i.e. the quartic potential for an inflaton with NMDC is equivalent to $\varphi^{4/3}$ monomial potential for an inflaton with minimal coupling. The quartic coupling $\lambda_4 \equiv \lambda_\phi$ is depicted to the dimensionful $\xi$-parameter  $\xi^{8/3}_\varphi= \lambda_\phi^{1/3}(3M_\text{Pl}\tilde{M})^{4/3}$.

Also, during de-Sitter phase, the quadratic potential $V\propto \phi^2$ with non-minimal kinetic coupling is dual with the potential
\begin{equation}
V(\phi)=\lambda_2 M^2_\text{Pl} \phi^2\equiv \frac12 m^2_\phi \phi^2 \quad \longleftrightarrow \quad V_\text{GR}(\varphi) = \lambda_2 M^2_\text{Pl} \left(2\lambda_2^{-1/2} \tilde{M} \right)\varphi \equiv  \mu^3_\varphi \varphi
\end{equation}
where, the mass squared $m^2_\phi\equiv 2\lambda_2 M^2_\text{Pl}$ is depicted to $\mu^3_\varphi=\sqrt{2}\,m_\phi \,\tilde{M}\,M_\text{Pl}$.
Furthermore, during the de-Sitter phase the linear potential $V(\phi) =\mu^3_\phi \phi$ with NMDC is dual with $V_\text{GR}(\varphi)=(3/2)^{2/3} \mu^2_\phi(\tilde{M} M_\text{Pl})^{2/3} \varphi^{2/3}$;  see also Ref. \cite{McAllister:2014mpa} for relevant monomial potentials in stringy set ups. For the case $n=-2$ the potential expression (\ref{vmm}) cannot be used. The inverse quadratic potential $V_\text{}(\phi)=m^6\phi^{-2}$ is instead depicted to an exponential potential.

We comment that for the case $n=-2$ the potential expression (\ref{vmm}) cannot be used. The inverse quadratic potential $V_\text{}(\phi)=m^6\phi^{-2}$ is instead depicted to an exponential potential, $V_\text{}(\phi)\propto {\phi^{-2}}  \longleftrightarrow V_\text{GR}(\varphi) \propto e^{-2\frac{M_\text{Pl}\tilde{M}}{m^3}\varphi}$ for $\phi, \varphi>0$.

It is interesting to note that if we define
\begin{equation} \label{vmm2}
V_\text{GR}(\varphi) = \lambda_pM^{4-p}_\text{Pl}\left(\frac{p+2}{2} \lambda^{-1/2}_p M_\text{Pl}^{p/2-1}\tilde{M}\,\varphi \right)^{2p/(p+2)}\, \equiv \, \lambda_q \, M^{4-q}_\text{Pl} \,\varphi^q
\end{equation}
then the CMB normalized $\lambda_q$ values, see Eq. (\ref{gener}), are automatically depicted to CMB normalized values $\lambda_p$ or $\lambda_e$ and the expressions (\ref{lp}) or (\ref{le}) are re-derived.

\subsection{The slow-roll parameters}

The well-known GR result for the spectral index $n_s$ for monomial potentials (\ref{A}) can be directly transformed, by the correspondence relation (\ref{corre}), into the NMDC result, see Eq. (\ref{A-nmdc}), namely $1-n_s\simeq 2(p+1)[(p+2) N]^{-1}$. On the other hand, the expressions for slow-roll parameters $\epsilon$ and $\eta$, derived by considering the dynamics of the NMDC inflaton, are given by Eq. (\ref{epet}), $\epsilon \simeq p [(2p+4) N]^{-1} $ and $\eta \simeq (p-1)[(p+2) N]^{-1}$. We can combine the two results, (\ref{A-nmdc}) and (\ref{epet}), and derive the NMDC-modified relation (\ref{modif}) for the spectral index in terms of the slow-roll parameters. By equating
\begin{equation}
1-n_s = \alpha \epsilon +\beta \eta
\end{equation}
we directly find that $\alpha=8$ and $\beta=-2$, in agreement with the (\ref{modif}).

\end{document}